\documentclass[aps,superscriptaddress,prd,
onecolumn,
floatfix,
nofootinbib,
longbibliography,
amsmath,amssymb,amsfonts]{revtex4-2}

\usepackage{graphicx}
\usepackage{dcolumn}
\usepackage{bm}

\usepackage[caption=false]{subfig}
\usepackage[colorlinks,linkcolor=blue,anchorcolor=violet,citecolor=red]{hyperref}

\usepackage{graphicx}
\usepackage{subfig}
\usepackage{mathrsfs}

\newcommand{\sm}{-}

\usepackage{orcidlink}
\begin{document}


\title{Squeezed quantum states and partner modes in the moving mirror model of black hole evaporation
}

\author{Kuan-Nan Lin\, \orcidlink{0000-0001-6561-4255}}
\email{knlinphy@gmail.com}
\affiliation{Asia Pacific Center for Theoretical Physics (APCTP), Pohang, 37673, Korea}

\author{Pisin Chen\, \orcidlink{0000-0001-5251-7210}}
\email{pisinchen@phys.ntu.edu.tw}
\affiliation{
Leung Center for Cosmology and Particle Astrophysics, National Taiwan University, Taipei 10617, Taiwan (R.O.C.)
}
\affiliation{Department of Physics and Center for Theoretical Sciences, National Taiwan University, Taipei 10617, Taiwan (R.O.C.)}
\affiliation{Graduate Institute of Astrophysics, National Taiwan University, Taipei 10617, Taiwan (R.O.C.)}
\affiliation{Kavli Institute for Particle Astrophysics and Cosmology,
SLAC National Accelerator Laboratory, Stanford University, Stanford, California 94305, USA}

\author{Michael R.R. Good\, \orcidlink{0000-0002-0460-1941}}
\email{muon@asu.edu}
\affiliation{
Leung Center for Cosmology and Particle Astrophysics, National Taiwan University, Taipei 10617, Taiwan (R.O.C.)
}
\affiliation{Physics Department \& Energetic Cosmos Laboratory, Nazarbayev University,
Astana 010000, Qazaqstan.}
\affiliation{Beyond Center for Fundamental Concepts in Science, Arizona State University,
Tempe AZ 85287, USA.}

\author{Yasusada Nambu\, \orcidlink{0000-0003-2596-4650}}
\email{yasusada.nambu@gmail.com}
\affiliation{Department of Physics, Nagoya University, Nagoya 464-8602, Japan}
\affiliation{%
Kansai Institute for Photon Science (KPSI), National Institute for Quantum Science and Technology (QST), Kizugawa,
 Kyoto 619-0215, Japan
}%
\thanks{Visiting Collaborative Researcher at KPSI,QST}


\date{\today}

\begin{abstract}
The standard moving mirror model in (1+1)-dimensional spacetime is known to reproduce several quantum aspects of black hole evaporation. A perfectly reflecting, accelerating mirror can emit radiation whose frequency spectrum resembles that of Hawking radiation, although this correspondence holds only under certain approximations. Moreover, the standard \textit{in-out} formulation does not provide a natural notion of the partner modes associated with the Hawking radiation. In this paper, we reformulate the moving mirror model in terms of Rindler/Milne modes. This formulation not only attributes the origin of the required approximations to mode squeezing effects but also naturally incorporates the notion of partner modes. Furthermore, as a consequence of these mode squeezing effects, the radiation received by an inertial observer at future null infinity exhibits additional nontrivial quantum correlations, even though its frequency spectrum approximately follows a Bose--Einstein or Fermi--Dirac distribution.

\end{abstract}

\maketitle

\tableofcontents

\section{Introduction}
\label{intro}

A semiclassical black hole is known to emit quantum radiation \cite{Hawking1975}. However, in the absence of a complete theory of quantum gravity, the proper-time evolution of the black hole beyond the semiclassical regime remains poorly understood. If the semiclassical approximation is extrapolated to the final stage of black hole evaporation, the entangled partner modes of the Hawking radiation are expected to fall into the black hole and eventually disappear. Consequently, the initial pure-state pairs of Hawking quanta and their entangled partners can no longer be fully recovered \cite{PhysRevD.14.2460}. This apparent loss of quantum information violates the unitarity of quantum mechanics and constitutes the well-known \textit{black hole information paradox}.

An alternative approach is provided by a perfectly reflecting, point-like mirror with prescribed motion in (1+1)-dimensional flat spacetime. Such a moving mirror is known to emit quantum radiation, referred to as \textit{mirror-induced radiation} (MIR), which can mimic key features of Hawking radiation from semiclassical black holes in higher-dimensional curved spacetimes \cite{Fulling1976,Davies1977}. By appropriately designing the mirror's trajectory, several black-hole evaporation scenarios have been proposed \cite{PhysRevD.36.2327,PTEPHotta,PhysRevD.96.025016,Good2017,PhysRevD.110.025023,Lin:2024ihr}, providing a useful framework for investigating the dynamics of evaporation and gaining insight into the black hole information problem.

The notion of particles in MIR is typically defined in terms of Minkowski plane wave modes associated with inertial observers to the mirror's right at the past $\mathscr{I}^{-}_{R}$ and future $\mathscr{I}^{+}_{R}$ null infinities, which define the \textit{in} and \textit{out} vacua, respectively. However, within this \textit{in-out} description, the notion of entanglement between the radiated particles is not manifest. In particular, the Minkowski plane wave \textit{out} modes alone already constitute a complete basis at $\mathscr{I}^{+}_{R}$. To characterize entanglement, the quantum field at $\mathscr{I}^{+}_{R}$ must instead be decomposed into two mode bases that together form a complete basis. Consequently, if one measures quantum correlations between particles defined with respect to the Minkowski plane wave \textit{out} modes in the laboratory, additional operations are required to reveal and quantify the underlying quantum entanglement.

From the perspective of the frequency spectrum of the radiated particles, as defined by the Minkowski plane wave \textit{out} modes, certain approximations \cite{Hawking1975,DeWitt1975,Davies1977,haro2005moving,Good:2011nue,PhysRevD.88.025023} are required to recover a Bose--Einstein distribution for bosons (or a Fermi--Dirac distribution for fermions). These thermal distributions are commonly regarded as a characteristic signature of analogue Hawking radiation. However, the physical interpretation and validity of these approximations remain unclear. In particular, it is not yet clear to what extent the resulting radiated particles can genuinely be interpreted as analogue Hawking radiation.

The aim of this paper is to provide answers to the questions:
\begin{itemize}
    \item \textit{Q1: What is the meaning of the approximations made to obtain a Bose--Einstein distribution?} 
    \item \textit{Q2: Where is the notion of quantum entanglement/partner modes in the mirror-induced radiation?} 
\end{itemize}

As we will demonstrate later, the brief answers to these questions are:
\begin{itemize}
    \item \textit{A1: The mirror-field interaction introduces additional squeezing effect. } 
    \item \textit{A2: The notion of entanglement can be extracted by considering the Rindler/Milne modes. } 
\end{itemize}

\begin{figure}[h]
    \centering
    \includegraphics[width=0.8\linewidth]{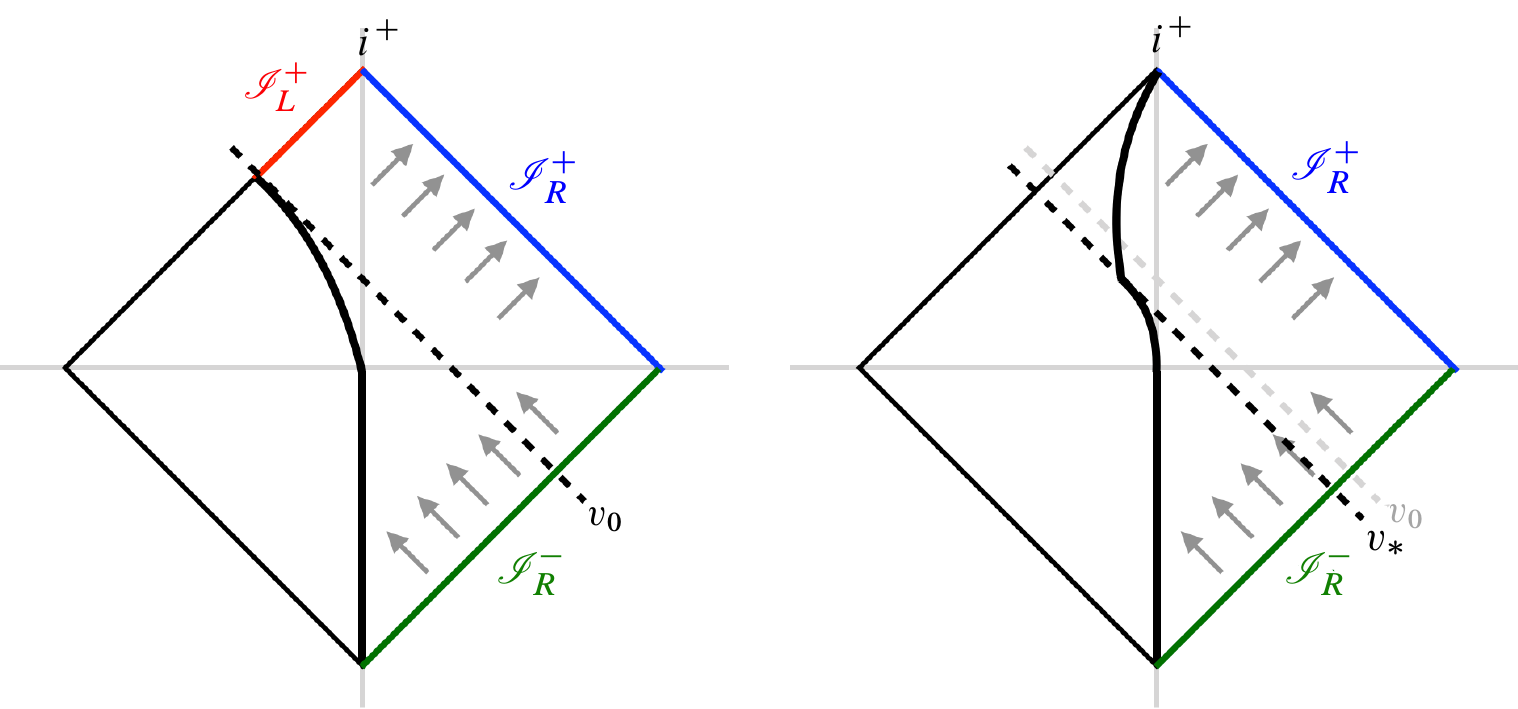}  
    \caption{ Penrose diagrams of moving mirror spacetimes. Left: A mirror (thick black) begins from the past timelike infinity and approaches the asymptote $v_{0}$. The past null infinity (green) is a Cauchy surface in the past, while the right future null infinity (blue) and the left future null infinity to the mirror's right (red) together form a Cauchy surface in the future. Right: Before reaching the asymptote $v_{0}$, the mirror is decelerated at $v_{*}<v_{0}$ and ends up at the future timelike infinity. In this case, the right future null infinity (blue) is a Cauchy surface in the future.  }
    \label{fig-penrose}
\end{figure}

If the mirror is asymptotically null, a Cauchy surface in the future may consist of both the right and left future null infinities on one side, say, right, of the moving mirror (see Fig.~\ref{fig-penrose}). In this case, the radiation arriving at $\mathscr{I}_{R}^{+}$ does not cover the entire Cauchy surface. On the contrary, if the mirror ends up at the future timelike infinity $i^{+}$, the radiation arriving at $\mathscr{I}_{R}^{+}$ now does cover the entire Cauchy surface. Depending on where in the future does the mirror ends up at, the content of the radiation observed by an inertial observer at $\mathscr{I}_{R}^{+}$ will be different.

In the case of asymptotically null motions, the radiation arriving at $\mathscr{I}_{R}^{+}$ may be naturally associated with Hawking radiation in some way, while the radiation at $v>v_{0}$ (see Fig.~\ref{fig-penrose}) that never interacts with the mirror, which prevents the inertial observer at $\mathscr{I}_{R}^{+}$ from accessing its degrees of freedom, is natural to be associated with the partner mode of the Hawking radiation. This mimics the trapping of the partner mode inside the horizon of a black hole.

In the case that the mirror ends up at $i^{+}$, both the Hawking mode and its partner mode will be reflected by the mirror and become accessible to the inertial observer at $\mathscr{I}_{R}^{+}$. In other words, the notion of particles defined by the inertial observer at $\mathscr{I}_{R}^{+}$ should be a mixture of both the Hawking mode and its partner mode. Therefore, the notion of entanglement between the Hawking mode and its partner mode emerges only if one partitions the Cauchy surface $\mathscr{I}_{R}^{+}$ into two subregion, with each subregion only having access to either the Hawking mode or the partner mode.

In particular, we provide answers to the above questions by employing the notions of Milne mode (Hawking mode) in $v<v_{0}$ and Rindler mode (partner mode) in $v>v_{0}$, which are the  standard entangled partner pairs of each other, separated by the asymptote $v=v_{0}$ \cite{PhysRevD.14.870,PhysRevD.100.065019}. Essentially, it turns out that in the case that a mirror travels to the left and eventually approaches asymptotically null, the right-moving particles radiated to the inertial observer at $\mathscr{I}_{R}^{+}$ is a two-mode squeezing
of the reflected Milne modes, which are initially emanated from $\mathscr{I}_{R}^{-}$ and reflected by the receding mirror, while the left-moving particles observed at $\mathscr{I}_{L}^{+}$ on the mirror's right-hand side is simply the left-moving Rindler modes, which are also emanated from $\mathscr{I}_{R}^{-}$ but never interacts with the mirror. Due to the squeezing, the right-moving particles that arrived at $\mathscr{I}_{R}^{+}$ do not have an exact Bose--Einstein spectrum, and they possess self-correlation in addition to the mutual-correlation with the left-moving Rindler modes. In contrast, the standard Rindler/Milne modes are known to possess exact Bose--Einstein spectra, and they are only mutually correlated without self-correlations.

In the case where the mirror returns to future timelike infinity, the right-moving particles observed by the inertial observer at $\mathscr{I}_{R}^{+}$ involve individual two-mode squeezing of the Rindler/Milne modes and mutual two-mode squeezing between them. The multiple squeezings distort the frequency spectrum of the right-moving particles observed by the inertial observer at $\mathscr{I}_{R}^{+}$ and complexify the correlations. In this paper, we will show how to extract the information of the Rindler/Milne modes from the right-moving radiation at $\mathscr{I}_{R}^{+}$ to develop physical insights.

This paper is organized as follows. In Sect.~\ref{Entanglement structure of vacuum}, we derive the vacuum structure in cases where the moving mirror spacetime either possesses an asymptote or does not. In Sect.~\ref{Correlation among the particles}, we review the standard derivation of the radiation correlation functions. In Sect.~\ref{Particle creation by perfect reflection}, the standard discussion of particle creation by a perfect mirror is reviewed. In Sect.~\ref{Formation of eternal black holes}, we consider the case in which the mirror mimics the formation of a black hole and provide a formulation in terms of the Rindler/Milne modes. In Sect.~\ref{Formation and evaporation of black holes}, we consider the case in which the mirror mimics the formation and subsequent evaporation of a black hole and provide a formulation in terms of the Rindler/Milne modes. Finally, in Sect.~\ref{Discussions}, we discuss the implications of the results and outline possible future research directions. Throughout this work, we use units with $c=\hbar=1$.


\section{Entanglement structure of vacuum}
\label{Entanglement structure of vacuum}

In this section, we investigate the structure of vacuum in the standard moving mirror model in (1+1)d Minkowski spacetime. To mimic Hawking radiation emitted by a black hole using a perfectly reflecting mirror in flat spacetime, the mirror must undergo a sufficiently long period of extreme acceleration such that the incident vacuum fluctuations are exponentially red-shifted forward in time (or, equivalently, blue-shifted backward in time) upon reflection. The Doppler effect experienced by the vacuum fluctuations in flat spacetime essentially plays the role of the gravitational effect in curved spacetimes. In addition, if the back reaction of the emission of Hawking radiation is negligible in reducing the mass of the black hole, then the corresponding moving mirror must be asymptotically null in the future so that the Hawking radiation can be continuously radiated at a given temperature, which is related to the black hole's mass (see Fig.~\ref{fig-penrose}). This belongs to the semi-classical regime of black hole evaporation.

Suppose the decrease in the mass of the black hole is to be taken into account; then there is not yet a satisfactory description of the evaporation process due to the lack of a widely accepted quantum theory of gravity. However, there have recently been proposals \cite{engel15,alm20,pen22,chen22,chen23} that attempt to resolve the so-called \textit{black hole information paradox}, which arises from extrapolating the semi-classical description of black hole evaporation to its very final stage. If one believes that the process of black hole evaporation is unitary, then one should explain how the partner modes are released outside the black hole before complete evaporation.

Although the quantum gravity mechanism that may trigger the release of the partner modes remains unclear at present, one may gain intuition from the moving mirror model. For example, it is typical to model the black hole's evaporation by requiring that the mirror eventually stop accelerating (see Fig.~\ref{fig-penrose}). In this case, there is no longer a true asymptotically null line that mimics the black hole event horizon; therefore, the partner mode may eventually get reflected by the mirror and become accessible to an observer away from the mirror.

In the following, we discuss separately the asymptotically null case and the case in which the mirror ends up at future timelike infinity, since the structures of the corresponding vacuum states are distinct.

\subsection{Timelike to asymptotically null}

Let us consider the case where a perfect mirror begins from the past timelike infinity and becomes asymptotically null in the future. A scalar field operator that is restricted to the mirror's right-hand side can be expanded by
\begin{equation}
    \begin{aligned}
        \hat{\phi}(t,x)
        &=
        \int_{0}^{\infty}d\omega' 
        \big[
        \hat{a}_{-\omega' }\, u_{-\omega' }(t,x)
        +
        \hat a_{-\omega'}^\dag\,\bar u_{-\omega'}(t,x)
        \big]
        \\
        &=
        \int_{0}^{\infty}d\omega
        \big[
        \hat{b}_{\omega}^{R}\, v_{\omega}^{R}(t,x)
        +
        \hat{b}_{-\omega}^{L}\, v_{-\omega}^{L}(t,x)
        +
        h.c.
        \big].
    \end{aligned}
\end{equation}
 Here, $u_{-\omega'}$ is the left-moving Minkowski plane-wave \textit{in} mode, characterized by the null coordinate $v = t + x \in (-\infty, +\infty)$ at the right past null infinity $\mathscr{I}_{R}^{-}$ and $\bar u_{-\omega'}$ represents the complex conjugate of $u_{-\omega'}$. In the second line, $v_{\omega}^{R}$ is the right-moving Minkowski plane-wave \textit{out} mode, associated with $u = t - x \in (-\infty, +\infty)$ at the right future null infinity $\mathscr{I}_{R}^{+}$, but with $v \in (-\infty, v_{0})$ at $\mathscr{I}_{R}^{-}$, where $v_{0}$ denotes the asymptote approached by the mirror. The function $v_{-\omega}^{L}$ is the “complemented” left-moving \textit{out} mode, which does not reflect off the mirror and therefore is not a Minkowski plane wave; it is supported on $v \in (v_{0}, +\infty)$ at both the left future null infinity $\mathscr{I}_{L}^{+}$ and the right past null infinity $\mathscr{I}_{R}^{-}$. Observe that the pair $\{v_{\omega}^{R}, v_{-\omega}^{L}\}$ form a complete set of mode bases on the right side of the perfect mirror at the future null infinities, whereas $u_{-\omega'}$ by itself provides a complete basis at $\mathscr{I}_{R}^{-}$. Moreover, the \textit{in} vacuum $|0;in\rangle$ is defined via the annihilation condition $\hat{a}_{-\omega'}|0;in\rangle = 0 \;\forall\; \omega' > 0$, while the operators $\{\hat{b}_{\omega}^{R}, \hat{b}_{-\omega}^{L}\}$ specify the \textit{out} vacua through $\hat{b}_{\omega}^{R}|0\rangle_{R} = \hat{b}_{-\omega}^{L}|0\rangle_{L} = 0 \;\forall\; \omega > 0$. Finally, the particle notion for an inertial observer at $\mathscr{I}_{R}^{+}$ is associated with excitations of the mode $\hat{b}_{\omega}^{R}$.

The mode functions and the operators have the following Bogoliubov transformations:
\begin{equation}\label{bogo-RL-in}
    \begin{aligned}
        v_{\omega}^{R}
        &=
        \int_{0}^{\infty}d\omega' 
        \big(
        \alpha_{\omega,\sm\omega'}^{R}
        u_{\sm\omega' }
        +
        \beta_{\omega,\sm\omega'}^{R}
        \bar{u}_{\sm\omega' }
        \big),
        \\
        v_{-\omega}^{L}
        &=
        \int_{0}^{\infty}d\omega' 
        \big(
        \alpha_{\sm\omega,\sm\omega'}^{L}
        u_{\sm\omega' }
        +
        \beta_{\sm\omega,\sm\omega'}^{L}
        \bar{u}_{\sm\omega' }
        \big),
        \\
        \hat{b}_{\omega}^{R}
        &=
        \int_{0}^{\infty}d\omega' 
        \big(
        \bar{\alpha}_{\omega,\sm\omega'}^{R}
        \hat{a}_{\sm\omega' }
        -
        \bar{\beta}_{\omega,\sm\omega'}^{R}
        \hat{a}_{\sm\omega' }^{\dagger}
        \big),
        \\
        \hat{b}_{-\omega}^{L}
        &=
        \int_{0}^{\infty}d\omega' 
        \big(
        \bar{\alpha}_{\sm\omega,\sm\omega'}^{L}
        \hat{a}_{\sm\omega' }
        -
        \bar{\beta}_{\sm\omega,\sm\omega'}^{L}
        \hat{a}_{\sm\omega' }^{\dagger}
        \big),
    \end{aligned}
\end{equation}
where $\{\alpha_{\omega,\sm\omega'}^{R},\beta_{\omega,\sm\omega'}^{R},\alpha_{\sm\omega,\sm\omega'}^{L},\beta_{\sm\omega,\sm\omega'}^{L}\}$ are the Bogoliubov coefficients.

Since the \textit{out} modes $\{\hat{b}_{\omega}^{R},\hat{b}_{-\omega}^{L}\}$ together span across a Cauchy surface, $|0;in\rangle$ should involve the tensor product of the local states defined by $\{\hat{b}_{\omega}^{R},\hat{b}_{-\omega}^{L}\}$. However, for general Bogoliubov coefficients $\{\alpha_{\omega,\sm\omega'}^{R},\beta_{\omega,\sm\omega'}^{R},\alpha_{\sm\omega,\sm\omega'}^{L},\beta_{\sm\omega,\sm\omega'}^{L}\}$, deriving the expression of the \textit{in-vacuum} state may be more technically complicated since the $R$- and $L$-Bogoliubov coefficients are not identical and non-diagonal. To circumvent this technical difficulty, let us expand the \textit{out} modes in terms of the Rindler/Milne modes\footnote{As far as the \textit{in} vacuum at $\mathscr{I}_{R}^{-}$ is the Minkowski vacuum, it can always be expanded as an entangled state defined by the Rindler/Milne modes at $\mathscr{I}_{R}^{-}$, regardless of the mirror's trajectory or the existence of the mirror. However, to make such an expansion, one is required to specify a constant null curve $v_{0}$, which is mathematically arbitrary, that separates the Rindler/Milne modes to each side of $v_{0}$. This is mathematically possible since, at $\mathscr{I}_{R}^{-}$, both the Rindler/Milne modes together form a complete bases, i.e., the Rindler mode is a complete basis on the one side of $v_{0}$ and the Milne mode is a complete basis on the other side of $v_{0}$. However, in order for this expansion to have a clear physical meaning, we consider the trajectories of the mirror such that there exist either an asymptotically or a would-be null curve, which provide a natural identification for $v_{0}$ and are equivalent to the time-dependent peeling function of the trajectory possessing a certain period of a constant plateau given by $v_{0}^{-1}$ \cite{Good:2019tnf,Ievlev:2022emz}.  } instead:
\begin{equation}\label{bogo-RL-I,II}
    \begin{aligned}
        v_{\omega}^{R}
        &=
        \int_{0}^{\infty}d\Omega' 
        \big(
        \alpha_{\omega,\sm\Omega'}^\text{II}
        u_{\sm\Omega' }^\text{II}
        +
        \beta_{\omega,\sm\Omega'}^\text{II}
        \bar{u}_{\sm\Omega' }^\text{II}
        \big),
        \\
        v_{-\omega}^{L}
        &=
        \int_{0}^{\infty}d\Omega' 
        \big(
        \alpha_{\sm\omega,\sm\Omega'}^\text{I}
        u_{\sm\Omega' }^\text{I}
        +
        \beta_{\sm\omega,\sm\Omega'}^\text{I}
        \bar{u}_{\sm\Omega' }^\text{I}
        \big),
        \\
        \hat{b}_{\omega}^{R}
        &=
        \int_{0}^{\infty}d\Omega' 
        \big(
        \bar{\alpha}_{\omega,-\Omega'}^\text{II}
        \hat{a}_{-\Omega' }^\text{II}
        -
        \bar{\beta}_{\omega,-\Omega'}^\text{II}
        (\hat{a}_{-\Omega' }^\text{II})^{\dagger}
        \big),
        \\
        \hat{b}_{-\omega}^{L}
        &=
        \int_{0}^{\infty}d\Omega' 
        \big(
        \bar{\alpha}_{-\omega,-\Omega'}^\text{I}
        \hat{a}_{-\Omega' }^\text{I}
        -
        \bar{\beta}_{-\omega,-\Omega'}^\text{I}
        (\hat{a}_{-\Omega' }^\text{I})^{\dagger}
        \big),
    \end{aligned}
\end{equation}
where $u_{-\Omega' }^\text{I}$ is the left-moving Rindler mode that covers $v\in(v_{0},\infty)$ at $\mathscr{I}_{R}^{-}$, $u_{-\Omega' }^\text{II}$ is the left-moving Milne mode that covers $v\in(-\infty,v_{0})$, and $\{\hat{a}_{-\Omega'}^\text{I},\hat{a}_{-\Omega'}^\text{II}\}$ are the corresponding annihilation operators. The inverse transformations are 
\begin{equation}\label{bogo-uI-II,vRL}
    \begin{aligned}
        u_{-\Omega'}^\text{I}
        &=
        \int_{0}^{\infty}d\omega
        \big(
        \bar{\alpha}_{-\omega,-\Omega'}^\text{I}
        v_{-\omega}^{L}
        -
        \beta_{-\omega,-\Omega'}^\text{I}
        \bar{v}_{-\omega}^{L}
        \big),
        \\
        u_{-\Omega'}^\text{II}
        &=
        \int_{0}^{\infty}d\omega
        \big(
        \bar{\alpha}_{\omega,-\Omega'}^\text{II}
        v_{\omega}^{R}
        -
        \beta_{\omega,-\Omega'}^\text{II}
        \bar{v}_{\omega}^{R}
        \big),
        \\
        \hat{a}_{-\Omega'}^\text{I}
        &=
        \int_{0}^{\infty}d\omega
        \big(
        \alpha_{-\omega,-\Omega'}^\text{I}
        \hat{b}_{-\omega}^{L}
        +
        \bar{\beta}_{-\omega,-\Omega'}^\text{I}
        (\hat{b}_{-\omega}^{L})^{\dagger}
        \big),
        \\
        \hat{a}_{-\Omega'}^\text{II}
        &=
        \int_{0}^{\infty}d\omega
        \big(
        \alpha_{\omega,-\Omega'}^\text{II}
        \hat{b}_{\omega}^{R}
        +
        \bar{\beta}_{\omega,-\Omega'}^\text{II}
        (\hat{b}_{\omega}^{R})^{\dagger}
        \big).
    \end{aligned}
\end{equation}

With the Rindler/Milne modes at hand, one may further construct the Unruh modes, which are obtained by the analytic continuation of the Rindler/Milne modes through the diagonal Bogoliubov transformations \cite{PhysRevD.14.870}: 
\begin{equation}
    \begin{aligned}
        h_{-\Omega'}^{(1)}
        &=
        \bar{\tilde{\alpha}}_{-\Omega'}u_{-\Omega'}^\text{I}
        -
        \tilde{\beta}_{-\Omega'}\bar{u}_{-\Omega'}^\text{II},
        \\
        h_{-\Omega'}^{(2)}
        &=
        \bar{\tilde{\alpha}}_{-\Omega'}u_{-\Omega'}^\text{II}
        -
        \tilde{\beta}_{-\Omega'}\bar{u}_{-\Omega'}^\text{I},
        \\
        \hat{a}_{-\Omega'}^{(1)}
        &=
        \tilde{\alpha}_{-\Omega'}\hat{a}_{-\Omega'}^\text{I}
        +
        \bar{\tilde{\beta}}_{-\Omega'}(\hat{a}_{-\Omega'}^\text{II})^{\dagger},
        \\
        \hat{a}_{-\Omega'}^{(2)}
        &=
        \tilde{\alpha}_{-\Omega'}\hat{a}_{-\Omega'}^\text{II}
        +
        \bar{\tilde{\beta}}_{-\Omega'}(\hat{a}_{-\Omega'}^\text{I})^{\dagger},
    \end{aligned}
\end{equation}
where
\begin{equation}
    \begin{aligned}
        \tilde{\alpha}_{-\Omega'}
        =
        \frac{e^{\pi\Omega'/2\kappa}}{\sqrt{2\sinh(\pi\Omega'/\kappa)}},
        \quad
        \tilde{\beta}_{-\Omega'}
        =
        -
        \frac{e^{-\pi\Omega'/2\kappa}}{\sqrt{2\sinh(\pi\Omega'/\kappa)}},
    \end{aligned}
\end{equation}
and $v_{0}=1/\kappa$ is the asymptote (horizon) separating the Rindler mode and the Milne mode.

By using $\hat{a}_{-\Omega'}^{(1)}|0;in\rangle=\hat{a}_{-\Omega'}^{(2)}|0;in\rangle=0$, $|0;in\rangle$ is solved as a mutual two-mode squeezed state:
\begin{equation}\label{invac-asym}
    \begin{aligned}
        |0;in\rangle
        &=
        \prod_{\Omega'>0}|0_{-\Omega'};in\rangle,
    \end{aligned}
\end{equation}
in terms of the Rindler/Milne modes, where
\begin{equation}\label{rindler/milne-entangle}
    \begin{aligned}
        |0_{-\Omega'};in\rangle
        &:=
        \frac{1}{|\tilde{\alpha}_{-\Omega'}|}\,
        \mathrm{exp}\left(-\frac{\bar{\tilde{\beta}}_{-\Omega'}}{\tilde{\alpha}_{-\Omega'}}(\hat{a}_{-\Omega'}
        ^\text{I})^{\dagger}(\hat{a}_{-\Omega'}^\text{II})^{\dagger}\right)
        |0\rangle_\text{I}|0\rangle_\text{II}
        \\
        &=
        \frac{1}{|\tilde{\alpha}_{-\Omega'}|}
        \sum_{n=0}^{\infty}
        \Bigl(
        -\frac{\bar{\tilde{\beta}}_{-\Omega'}}{\tilde{\alpha}_{-\Omega'}}
        \Bigr)^{n}
        |n_{-\Omega'}\rangle_\text{I}|n_{-\Omega'}\rangle_\text{II},
    \end{aligned}
\end{equation}
indicating that the Rindler mode and the Milne mode are an entangled pair with the two-mode squeezing parameter $-\bar{\tilde{\beta}}_{-\Omega'}/\tilde{\alpha}_{-\Omega'}$ in the \textit{in} vacuum, and they have identical frequencies $\{\Omega',\Omega'\}$ and momenta $\{-\Omega',-\Omega'\}$.

Now we would like to know how the vacua $|0\rangle_{R,L}$ defined by the \textit{out} modes $\{\hat{b}_{\omega}^{R},\hat{b}_{-\omega}^{L}\}$ are related to the \textit{in} vacuum. According to Eq.~\eqref{bogo-RL-I,II}, or Eq.~\eqref{bogo-uI-II,vRL}, there is a one-to-one mapping between the annihilation/creation operators of the \textit{out} modes and the Rindler/Milne modes, so one can quote the standard result of individual two-mode squeezing of the same particle type but different frequencies:\footnote{In contrast, Eq.~\eqref{rindler/milne-entangle} is a two-mode squeezed state of different particle types, i.e., Rindler and Milne, but identical frequency. We will name this as mutual two-mode squeezing in order to make distinction with the individual two-mode squeezing.}
\begin{equation}
    \begin{aligned}
        |0\rangle_\text{I}
        &=
        e^{iW_{I}}\hat{f}_\text{I}|0\rangle_{L},
        \\
        |0\rangle_\text{II}
        &=
        e^{iW_\text{II}}\hat{f}_\text{II}|0\rangle_{R},
    \end{aligned}
\end{equation}
where $e^{iW_\text{I}}={}_{L}\langle 0|0\rangle_\text{I},e^{iW_\text{II}}={}_{R}\langle 0|0\rangle_\text{II}$, and
\begin{equation}\label{fIfII-null}
    \begin{aligned}
        \hat{f}_\text{I}
        &=
        \mathrm{exp}
        \Big(
        -\frac{1}{2}
        \int_{0}^{\infty}d\omega_{1} d\omega_{2}\int_{0}^{\infty}d\Omega'
        (\alpha^{I})^{-1}_{-\Omega',-\omega_{2}}\bar{\beta}^\text{I}_{-\omega_{1},-\Omega'}
        (\hat{b}_{-\omega_{1}}^{L})^{\dagger}
        (\hat{b}_{-\omega_{2}}^{L})^{\dagger}
        \Big),
        \\
        \hat{f}_\text{II}
        &=
        \mathrm{exp}
        \Big(
        -\frac{1}{2}
        \int_{0}^{\infty}d\omega_{1} d\omega_{2}\int_{0}^{\infty}d\Omega'
        (\alpha^\text{II})^{-1}_{-\Omega',\omega_{2}}\bar{\beta}^\text{II}_{\omega_{1},-\Omega'}
        (\hat{b}_{\omega_{1}}^{R})^{\dagger}
        (\hat{b}_{\omega_{2}}^{R})^{\dagger}
        \Big).
    \end{aligned}
\end{equation}

By using Eq.~\eqref{bogo-uI-II,vRL} and the above equations, $|0;in\rangle$ can be expressed in terms of the original \textit{out} modes $\{\hat{b}_{\omega}^{R},\hat{b}_{-\omega}^{L}\}$ and vacua $\{|0\rangle_{R},|0\rangle_{L}\}$. The bottom line here is that the vacua $\{|0\rangle_\text{I},|0\rangle_\text{II}\}$ generally involve two-mode squeezing of the \textit{out} modes $\{\hat{b}_{\omega}^{R},\hat{b}_{-\omega}^{L}\}$ via the Bogoliubov coefficients: $\{\alpha^\text{I},\beta^\text{I},\alpha^\text{II},\beta^\text{II}\}$ and vise versa, i.e., $\{|0\rangle_{R},|0\rangle_{L}\}$ involve two-mode squeezing 
of the Rindler/Milne modes $\{\hat{a}_{-\Omega'}^\text{I},\hat{a}_{-\Omega'}^\text{II}\}$.

\subsection{Timelike to timelike}

If a black hole evaporates its mass away, then there will no longer be an event horizon in the spacetime. Therefore, the corresponding moving mirror should end up at the future timelike infinity instead of approaching asymptotically null. In this case, the field operator on the right-hand side of a perfectly reflecting mirror can be expanded solely in terms of the right-moving Minkowski plane wave \textit{out} mode $v_{\omega}^{R}$ at $\mathscr{I}_{R}^{+}$, and the left-moving Minkowski plane wave \textit{in} mode $u_{-\omega'}$ at $\mathscr{I}_{R}^{-}$ by 
\begin{equation}
    \begin{aligned}
        \hat{\phi}(t,x)
        &=
        \int_{0}^{\infty}d\omega' 
        \big[
        \hat{a}_{-\omega' }\, u_{-\omega' }(t,x)
        +
        h.c.
        \big]
        \\
        &=
        \int_{0}^{\infty}d\omega
        \big[
        \hat{b}_{\omega}^{R}\, v_{\omega}^{R}(t,x)
        +
        h.c.
        \big].
    \end{aligned}
\end{equation}

The Bogoliubov transformations between the \textit{in} and \textit{out} Fock spaces are thus simpler:
\begin{equation}
    \begin{aligned}
        v_{\omega}^{R}
        &=
        \int_{0}^{\infty}d\omega' 
        \big(
        \alpha_{\omega,-\omega'}^{R}
        u_{-\omega' }
        +
        \beta_{\omega,-\omega'}^{R}
        \bar{u}_{-\omega' }
        \big),
        \\
        \hat{b}_{\omega}^{R}
        &=
        \int_{0}^{\infty}d\omega' 
        \big(
        \bar{\alpha}_{\omega,-\omega'}^{R}
        \hat{a}_{-\omega' }
        -
        \bar{\beta}_{\omega,-\omega'}^{R}
        \hat{a}_{-\omega' }^{\dagger}
        \big),
        \\
        u_{-\omega' }
        &=
        \int_{0}^{\infty}d\omega 
        \big(
        \bar{\alpha}_{\omega,-\omega'}^{R}
        v_{\omega}^{R}
        -
        \beta_{\omega,-\omega'}^{R}
        \bar{v}_{\omega}^{R}
        \big),
        \\
        \hat{a}_{-\omega' }
        &=
        \int_{0}^{\infty}d\omega 
        \big(
        \alpha_{\omega,-\omega'}^{R}
        \hat{b}_{\omega}^{R}
        +
        \bar{\beta}_{\omega,-\omega'}^{R}
        (\hat{b}_{\omega}^{R})^{\dagger}
        \big),
    \end{aligned}
\end{equation}
and the \textit{in} vacuum is given by
\begin{equation}\label{invac-nonasym}
    \begin{aligned}
        |0;in\rangle
        &=
        e^{iW}\hat{f}|0;out\rangle,
    \end{aligned}
\end{equation}
where
\begin{equation}
    \begin{aligned}
        \hat{f}
        &=
        \mathrm{exp}
        \Big(
        -\frac{1}{2}
        \int_{0}^{\infty}d\omega_{1} d\omega_{2}
        \int_{0}^{\infty}d\omega'
        (\alpha^{R})^{-1}_{-\omega',\omega_{2}}\bar{\beta}^{R}_{\omega_{1},-\omega'}
        (\hat{b}_{\omega_{1}}^{R})^{\dagger}
        (\hat{b}_{\omega_{2}}^{R})^{\dagger}
        \Big),
    \end{aligned}
\end{equation}
and
\begin{equation}
    \begin{aligned}
        \int_{0}^{\infty}d\omega'
        \alpha^{R}_{\omega_{1},-\omega'}(\alpha^{R})^{-1}_{-\omega',\omega_{2}}
        =
        \delta(\omega_{1}-\omega_{2}).
    \end{aligned}
\end{equation}
In this case, the \textit{in} vacuum is a two-mode squeezed state of the right-moving Minkowski plane wave \textit{out} modes $\hat{b}_{\omega_{1,2}}^{R}$.

\section{Correlations among the particles}
\label{Correlation among the particles}

In principle, the structure of the vacuum states that we obtained in the previous section reveals the entanglement structure of the created particles in the \textit{in} vacuum state. However, the complex structures, i.e., multiple Bogoliubov transformations and integrals, make them not very practical for performing quantitative analysis. Therefore, in this paper, we shall instead focus on the correlation between various notions of particles defined by the number operator $\hat{N}_{k}$ of the specified mode \cite{fabbri2005modeling}: 
\begin{equation}
    \begin{aligned}
        Q(k_{1},k_{2})
        &:=
        \langle 0;in |\hat{N}_{k_{1}}\hat{N}_{k_{2}} | 0;in \rangle
        -
        \langle 0;in |\hat{N}_{k_{1}} | 0;in \rangle
        \langle 0;in |\hat{N}_{k_{2}} | 0;in \rangle,
        \quad
        k_{2}\neq k_{1},
    \end{aligned}
\end{equation}
where $\hat{N}_{k}$ is the number operator in mode $k\in(-\infty,\infty)$ with momentum $k$ and frequency $\omega=|k|$. 
%
The interpretation of this quantity is as follows: if $Q(k_{1},k_{2})=0$, the particle numbers in different modes $k_{1}$ and $k_{2}$ are not correlated; otherwise, they are correlated. For a thermofield double state, $Q(k_{1},k_{2})=0$ for $k_{1}k_{2}>0$ is well-known. Thus, this quantity is a kind  of the quantum correlation measure. This quantity is effectively identical to the second-order coherence, or intensity correlation, familiar from quantum optics. As is well established, the Hanbury-Brown–Twiss (HBT) effect \cite{Hanbury-Brown-Twiss-1954,Gardiner2004} relies on the properties of this intensity correlation and exposes the quantum character of light. In the context of the moving mirror model, the quantum correlations of analog Hawking radiation are analyzed using this same correlation function \cite{Tomonaga2024}.

\subsection{Timelike to asymptotically null}

In the case that the mirror is asymptotically null, say, to the left in the future, and suppose we only focus on the spacetime region to the right of the mirror, then there are three types of correlations of interests:
\begin{equation}
    \begin{aligned}
        Q^{RR}(k_{1},k_{2})
        &:=
        \langle 0;in |\hat{N}_{k_{1}}^{R}\hat{N}_{k_{2}}^{R} | 0;in \rangle
        -
        \langle 0;in |\hat{N}_{k_{1}}^{R} | 0;in \rangle
        \langle 0;in |\hat{N}_{k_{2}}^{R} | 0;in \rangle,
        \\
        Q^{LL}(k_{1},k_{2})
        &:=
        \langle 0;in |\hat{N}_{k_{1}}^{L}\hat{N}_{k_{2}}^{L} | 0;in \rangle
        -
        \langle 0;in |\hat{N}_{k_{1}}^{L} | 0;in \rangle
        \langle 0;in |\hat{N}_{k_{2}}^{L} | 0;in \rangle,
        \\
        Q^{RL}(k_{1},k_{2})
        &:=
        \langle 0;in |\hat{N}_{k_{1}}^{R}\hat{N}_{k_{2}}^{L} | 0;in \rangle
        -
        \langle 0;in |\hat{N}_{k_{1}}^{R} | 0;in \rangle
        \langle 0;in |\hat{N}_{k_{2}}^{L} | 0;in \rangle,
    \end{aligned}
\end{equation}
where $\hat{N}_{k}^{R,L}:=(\hat{b}_{k}^{R,L})^{\dagger}\,\hat{b}_{k}^{R,L}$ are the number operators defined by the \textit{out} modes $\{\hat{b}_{k}^{R},\hat{b}_{k}^{L}\}$. $Q^{RR}$ is the correlation between two right-moving \textit{out} particles radiated by the mirror, $Q^{LL}$ is the correlation between two left-moving \textit{out} particles
, and $Q^{RL}$ is the correlation between one right-moving \textit{out} particle and one left-moving \textit{out} particle. In terms of the Bogoliubov coefficients, one has
\begin{equation}\label{Q-perfect-asy}
    \begin{aligned}
        &Q^{RR}(k_{1}=\omega_{1}>0,k_{2}=\omega_{2}>0)
        =
        \left|
        \int_{0}^{\infty}d\omega_{1}' \beta_{k_{1},-\omega_{1}'}^{R}\bar{\beta}_{k_{2},-\omega_{1}'}^{R}
        \right|^{2}
        +
        \left|
        \int_{0}^{\infty}d\omega_{1}' \beta_{k_{1},-\omega_{1}'}^{R}\alpha_{k_{2},-\omega_{1}'}^{R}
        \right|^{2},
        \quad
        \omega_{1}\neq\omega_{2},
        \\
        &Q^{LL}(k_{1}=-\omega_{1}<0,k_{2}=-\omega_{2}<0)
        =
        \left|
        \int_{0}^{\infty}d\omega_{1}' \beta_{k_{1},-\omega_{1}'}^{L}\bar{\beta}_{k_{2},-\omega_{1}'}^{L}
        \right|^{2}
        +
        \left|
       \int_{0}^{\infty}d\omega_{1}' \beta_{k_{1},-\omega_{1}'}^{L}\alpha_{k_{2},-\omega_{1}'}^{L}
        \right|^{2},
        \quad
        \omega_{1}\neq\omega_{2},
        \\
        &Q^{RL}(k_{1}=\omega_{1}>0,k_{2}=-\omega_{2}<0)
        \\
        &=
        \int_{0}^{\infty}d\omega_{1}' \beta_{k_{1},-\omega_{1}'}^{R}\bar{\beta}_{k_{2},-\omega_{1}'}^{L}
        \int_{0}^{\infty}d\omega_{2}' \bar{\alpha}_{k_{1},-\omega_{2}'}^{R}\alpha_{k_{2},-\omega_{2}'}^{L}
        +
        \int_{0}^{\infty}d\omega_{1}' \beta_{k_{1},-\omega_{1}'}^{R}\alpha_{k_{2},-\omega_{1}'}^{L}
        \int_{0}^{\infty}d\omega_{2}' \bar{\alpha}_{k_{1},-\omega_{2}'}^{R}\bar{\beta}_{k_{2},-\omega_{2}'}^{L}.
    \end{aligned}
\end{equation}

On the one hand, since $Q^{RR}$ only involves the right-moving \textit{out} mode $\hat{b}^{R}$, which can be expressed purely in terms of the left-moving Milne mode $\hat{a}^{\mathrm{II}}$, there is no information about the entanglement between the Milne mode $\hat{a}^{\mathrm{II}}$ and the Rindler mode $\hat{a}^{\mathrm{I}}$, and vice versa for $Q^{LL}$. Therefore, $Q^{RR}\neq 0$ and $Q^{LL}\neq 0$ are simply the additional quantum correlations induced by the interaction of each mode with the mirror. On the other hand, $Q^{RL}$ involves both $\hat{b}^{R}$ and $\hat{b}^{L}$; thus, entanglement between $\hat{a}^{\mathrm{I}}$ and $\hat{a}^{\mathrm{II}}$ and their additional mirror-induced correlation are all encoded in $Q^{RL}$.

\subsection{Timelike to timelike}

In the case that the perfect mirror begins from the past timelike infinity and ends at the future timelike infinity, there is only one correlation to be studied. The correlation between two \textit{out} particles (defined by $\hat{b}_{k}^{R}$) radiated to the right-hand side of the perfect mirror is
\begin{equation}\label{Q-perfect-non-asy}
    \begin{aligned}
        Q^{RR}(k_{1}=\omega_{1}>0,k_{2}=\omega_{2}>0)
        =
        \left|
        \int_{0}^{\infty}d\omega_{1}' \beta_{k_{1},-\omega_{1}'}^{R}\bar{\beta}_{k_{2},-\omega_{1}'}^{R}
        \right|^{2}
        +
        \left|
        \int_{0}^{\infty}d\omega_{1}' \beta_{k_{1},-\omega_{1}'}^{R}\alpha_{k_{2},-\omega_{1}'}^{R}
        \right|^{2},
        \quad
        \omega_{1}\neq\omega_{2}.
    \end{aligned}
\end{equation}

In contrast to the previous case, here $Q^{RR}$ is the only measure of the correlation, and therefore the entanglement between the Rindler/Milne modes and their additional mirror-induced correlation are all encoded in $Q^{RR}$.

\section{Particle creation by perfect reflection}
\label{Particle creation by perfect reflection}

In this section, we briefly review the common procedure of computing the Bogoliubov coefficients: $\{\alpha_{\omega,-\omega'}^{R},\beta_{\omega,-\omega'}^{R}\}$, which are the components relevant to the reflection by the moving mirror. In addition, let us explicitly express the mode functions in terms of their momenta. Note that the discussions in this section are valid either the mirror is asymptotically null or ends up at the future timelike infinity.

For a perfectly reflecting mirror that begins from the past timelike infinity, the left-moving Minkowski plane wave \textit{in} mode $(p<0)$ and the right-moving Minkowski plane wave \textit{out} mode $(k>0)$ on the right-hand side of the mirror are related by
\begin{equation}
    v_{k}^{R}
    =
    \int_{-\infty}^{0} dp
    \big[
    \alpha_{kp}^{R}\,u_{p}
    +
    \beta_{kp}^{R}\,\bar{u}_{p}
    \big],
    \quad
    k>0,
\end{equation}
where $u_{p}=e^{-i\omega' v}/\sqrt{4\pi\omega'}$, $\omega'=-p>0$, at the past null infinity $\mathscr{I}^{-}_{R}$, and $v_{k}^{R}=e^{-i\omega u}/\sqrt{4\pi\omega}$, $\omega=k>0$, at the future null infinity $\mathscr{I}^{+}_{R}$, the annihilation/creation operators are related by
\begin{equation}\label{bogo-b-perfect}
    \hat{b}_{k}^{R}
    =
    \int_{-\infty}^{0} dp
    \big[
    \bar{\alpha}_{kp}^{R}\,\hat{a}_{p}
    -
    \bar{\beta}_{kp}^{R}\,\hat{a}_{p}^{\dagger}
    \big],
    \quad
    k>0.
\end{equation}

The Bogoliubov coefficients can be found by expanding the reflected Minkowski plane wave \textit{out} mode at the past null infinity $\mathscr{I}^{-}_{R}$, which is no longer a plane wave in terms of the Minkowski coordinates $(t,x)$, by
\begin{equation}
    \begin{aligned}
        -\frac{e^{-i\omega F(v)}}{\sqrt{4\pi\omega}}
        &=
        \int_{-\infty}^{0}dp
        \Big(
        \alpha_{\omega p}^{R}\frac{e^{-i\omega't+ipx}}{\sqrt{4\pi\omega'}}
        +
        \beta_{\omega p}^{R}\frac{e^{+i\omega't-ipx}}{\sqrt{4\pi\omega'}}
        \Big)
        \\
        &=
        \int_{0}^{+\infty}d\omega'
        \Big(
        \alpha_{\omega,-\omega'}^{R}\frac{e^{-i\omega' v}}{\sqrt{4\pi\omega'}}
        +
        \beta_{\omega,-\omega'}^{R}\frac{e^{+i\omega' v}}{\sqrt{4\pi\omega'}}
        \Big),
    \end{aligned}
\end{equation}
where $F(v)=T(v)-Z(T(v))$ is the ray-tracing function determined by the mirror's trajectory $(t,x)=(T,Z(T))$, which satisfies $v=t+x=T(v)+Z(T(v))$. If the mirror is asymptotically null to $v=v_{0}$ in the future, then $e^{-i\omega F(v)}\rightarrow \Theta(v_{0}-v)e^{-i\omega F(v)}$, where $\Theta$ is the Heaviside step function. The case where the mirror ends up at the future timelike infinity is equivalent to setting $v_{0}\rightarrow \infty$.

Equivalently, the reflected Minkowski plane wave \textit{in} mode may be expanded by the Minkowski plane wave \textit{out} mode at the future null infinity $\mathscr{I}^{+}_{R}$ by
\begin{equation}
    \begin{aligned}
        -\frac{e^{-i\omega' P(u)}}{\sqrt{4\pi\omega'}}
        &=
        \int^{+\infty}_{0}dk
        \Big(
        \bar{\alpha}_{k,-\omega'}^{R}\frac{e^{-i\omega t+ikx}}{\sqrt{4\pi\omega}}
        -
        \beta_{k,-\omega'}^{R}\frac{e^{+i\omega t-ikx}}{\sqrt{4\pi\omega}}
        \Big)
        \\
        &=
        \int_{0}^{\infty}d\omega
        \Big(
        \bar{\alpha}_{\omega,-\omega'}^{R}\frac{e^{-i\omega u}}{\sqrt{4\pi\omega}}
        -
        \beta_{\omega,-\omega'}^{R}\frac{e^{+i\omega u}}{\sqrt{4\pi\omega}}
        \Big),
    \end{aligned}
\end{equation}
where $P(u)=T(u)+Z(T(u))$ is the ray-tracing function determined by the mirror's trajectory $(t,x)=(T,Z(T))$ that satisfies $u=t-x=T(u)-Z(T(u))$.

From these expressions, one notices that the Bogoliubov coefficients are simply Fourier transforms:
\begin{equation}\label{bogo-conventional}
    \begin{aligned}
        \alpha_{\omega, -\omega'}^{R}
        &=
        -
        \frac{1}{2\pi}\sqrt{\frac{\omega'}{\omega}}
        \int_{-\infty}^{v_{0}}dv\, e^{i\omega' v}e^{-i\omega F(v)},
        \\
        \beta_{\omega, -\omega'}^{R}
        &=
        -
        \frac{1}{2\pi}\sqrt{\frac{\omega'}{\omega}}
        \int_{-\infty}^{v_{0}}dv\, e^{-i\omega' v}e^{-i\omega F(v)},
    \end{aligned}
\end{equation}
or
\begin{equation}
    \begin{aligned}
        \alpha_{\omega,-\omega'}^{R}
        &=
        -
        \frac{1}{2\pi}\sqrt{\frac{\omega}{\omega'}}
        \int_{-\infty}^{\infty}du\, e^{-i\omega u}e^{i\omega' P(u)},
        \\
        \beta_{\omega,-\omega'}^{R}
        &=
        \frac{1}{2\pi}\sqrt{\frac{\omega}{\omega'}}
        \int_{-\infty}^{\infty}du\, e^{-i\omega u}e^{-i\omega' P(u)}.
    \end{aligned}
\end{equation}

The equivalence of the two expressions can be verified by noting that the trajectory of the mirror in the lightcone coordinates is either $v=P(u)$ or $u=F(v)$, which implies $\{F,P\}$ are the inverses of each other since $F(P(u))=u$ and $P(F(v))=v$. Finally, in terms of the mirror's trajectory, one obtains
\begin{equation}\label{bogo-perfect-Z}
    \begin{aligned}
        \alpha_{\omega, -\omega'}^{R}
        &=
        -
        \frac{1}{2\pi}\sqrt{\frac{\omega'}{\omega}}
        \int_{-\infty}^{T_{0}}dT \big(1+Z'(T)\big)
        e^{-i(\omega-\omega')T+i(\omega+\omega')Z(T) },
        \\
        \beta_{\omega, -\omega'}^{R}
        &=
        -
        \frac{1}{2\pi}\sqrt{\frac{\omega'}{\omega}}
        \int_{-\infty}^{T_{0}}dT \big(1+Z'(T)\big)
        e^{-i(\omega+\omega')T+i(\omega-\omega')Z(T) },
    \end{aligned}
\end{equation}
where $T_{0}=[v_{0}+F(v_{0})]/2\rightarrow \infty$ for either asymptotically null or future timelike trajectories.

\section{Formation of eternal black holes}
\label{Formation of eternal black holes}

\subsection{Conventional Bose--Einstein approximation}
\label{Conventional Bose-Einstein approximation}

To mimic the emission of Hawking radiation during the gravitational collapse and formation of a black hole, one may employ the following mirror trajectory as an example \cite{Lin2020,Lin2021}:
\begin{equation}\label{traj-1}
    \begin{aligned}
        Z(T)
        =
        \begin{cases}
            0,
            \quad
            T\leq 0
            \\
            -T+\frac{1}{\kappa}-\frac{1}{\kappa}W\big(e^{1-2\kappa T}\big),
        \quad
        T \geq 0,
        \end{cases}
    \end{aligned}
\end{equation}
where $\kappa$ is a parameter of the trajectory related to the acceleration, and $W(x)$ is the Lambert W function, which is a solution to $We^{W}=x$ and has the expansion: $W(x)\simeq x$ for $x\ll 1$ recovering the Davies-Fulling motion \cite{Davies1977} at late times in the sense of $\kappa T\gg 1$, which represents the final formation of a black hole horizon.

According to Eq.~\eqref{bogo-perfect-Z}, the Bogoliubov coefficients involve integrations over the entire trajectory. However, if one extends the acceleration period to $T\leq 0$ by hand, a Bose--Einstein distribution can be obtained for a perfect mirror:
\begin{equation}\label{bogo-ideal}
    \begin{aligned}
        \beta_{\omega, -\omega'}^{R}
        &=
        -
        \frac{1}{2\pi}\sqrt{\frac{\omega'}{\omega}}
        \int_{-\infty}^{\infty}dT \big(1+Z'(T)\big)
        e^{-i(\omega+\omega')T+i(\omega-\omega')Z(T) }
        \\
        &=
        -
        \frac{ie^{-\frac{i\omega'}{\kappa}}}{2\pi \omega'}\sqrt{\frac{\omega'}{\omega}}e^{-\frac{\pi\omega}{2\kappa}}\big(\frac{\kappa}{\omega'}\big)^{\frac{i\omega}{\kappa}}
        \Gamma\big(1+\frac{i\omega}{\kappa}\big),
        \\
        \alpha_{\omega, -\omega'}^{R}
        &=
        -
        \frac{1}{2\pi}\sqrt{\frac{\omega'}{\omega}}
        \int_{-\infty}^{\infty}dT \big(1+Z'(T)\big)
        e^{-i(\omega-\omega')T+i(\omega+\omega')Z(T) }
        \\
        &=
        \frac{ie^{\frac{i\omega'}{\kappa}}}{2\pi \omega'}\sqrt{\frac{\omega'}{\omega}}e^{\frac{\pi\omega}{2\kappa}}\big(\frac{\kappa}{\omega'}\big)^{\frac{i\omega}{\kappa}}
        \Gamma\big(1+\frac{i\omega}{\kappa}\big),
    \end{aligned}
\end{equation}
which gives the desired spectrum that is usually believed to be an indication of (analog) Hawking radiation:
\begin{equation}\label{spec-hawking-standard}
    \begin{aligned}
        |\beta_{\omega, -\omega'}^{R}|^{2}
        &=
        \frac{1}{2\pi\kappa \omega'}\frac{1}{e^{\omega/T_{H}}-1},
    \end{aligned}
\end{equation}
where $T_{H}:=\kappa/2\pi$ is identified as the (analog) Hawking temperature. The extension of the accelerated motion to the infinite past, which corresponds to the Carlitz-Willey trajectory \cite{PhysRevD.36.2327}, \textit{by hand} is often argued to be a valid approximation for the late-time radiation spectrum \cite{DeWitt1975}, even though the Bogoliubov coefficients are time-independent. It is also known that the Bogoliubov coefficients derived from Eq.~\eqref{bogo-ideal} indicate that the radiated particles in different modes (i.e., momenta) are uncorrelated \cite{fabbri2005modeling} (see the discussion at the end of this subsection). Therefore, it is often said that the Hawking radiation from different modes is independent of one another. However, we will show that the exact, i.e., without extending the integration domain by hand, Bogoliubov coefficients not only give a distinct spectrum, but they also reveal non-vanishing correlations due to mode squeezing in the subsequent subsections \ref{Single-mode squeezing of Milne modes} and \ref{Additional correlations}.

Furthermore, Eq.~\eqref{spec-hawking-standard} is the spectrum radiated to the mirror's right-hand side due to the reflection by the receding mirror. Since the mirror is asymptotically null, one also has to take into consideration the left-moving modes that do not interact with the mirror, i.e., $\{\alpha_{-\omega, -\omega'}^{L},\beta_{-\omega, -\omega'}^{L}\}$. In contrast to the right-moving Minkowski plane wave \textit{out} mode $v_{\omega}^{R}$, which is a complete basis at $\mathscr{I}_{R}^{+}$, a left-moving Minkowski plane wave is over-complete at $\mathscr{I}_{L}^{+}$ on the right-hand side of the mirror. According to Wald \cite{wald1994quantum}, since a left-moving Rindler mode is complete at $v\in(v_{0},\infty)$, it may be used as the left-moving \textit{out} mode $v_{-\omega}^{L}$, with which one may compute $\{\alpha_{-\omega, -\omega'}^{L},\beta_{-\omega, -\omega'}^{L}\}$ explicitly.

To obtain the expression of $v_{-\omega}^{L}$, it is convenient to work in the lightcone coordinates $(u,v)$. The mirror's trajectory \eqref{traj-1} gives the ray-tracing function \cite{PhysRevD.80.125003}:
\begin{equation}\label{raytracing-1-F}
    \begin{aligned}
        F(v)
        &=
        \begin{cases}
            v,
            \quad
            v\leq 0
            \\
            -\frac{1}{\kappa}\ln(1-\kappa v),
            \quad
            0\leq v <v_{0}:=\frac{1}{\kappa}.
        \end{cases}
    \end{aligned}
\end{equation}

\begin{figure}[h]
    \centering
    \includegraphics[width=0.5\linewidth]{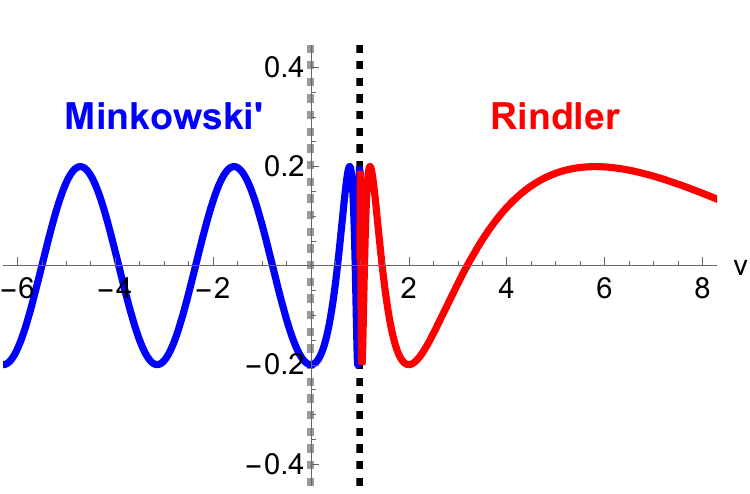}  
    \caption{Spacetime profiles of the Minkowski plane wave \textit{out} mode after reflection (denoted by a prime) and the Rindler \textit{out} mode at the past null infinity by taking the real parts of Eq.~\eqref{out Minkowski after reflection} and Eq.~\eqref{rindler-IR-}. The vertical black dashed line indicates the asymptote $v_{0}=1/\kappa$ (Rindler horizon), and the vertical gray dashed line is the moment the mirror begins to accelerate. We set $\kappa=1,\omega=2$.}
    \label{fig-profile1}
\end{figure}

Upon reflection backward in time, the right-moving Minkowski plane wave \textit{out} mode that is initially at $\mathscr{I}_R^+$ 
becomes, at $\mathscr{I}_R^-$, (see Fig.~\ref{fig-profile1})
\begin{equation}\label{out Minkowski after reflection}
    \begin{aligned}
        \left.v_{\omega}^{R}\right|_{\mathscr{I}_{R}^{-}}
        &=
        \begin{cases}
            -
        \frac{1}{\sqrt{4\pi\omega}}e^{-i\omega v},\quad v\leq 0
            \\
            -
        \frac{1}{\sqrt{4\pi\omega}}e^{\frac{i\omega}{\kappa}\ln(1-\kappa v)},
        \quad
        0\leq v< v_{0},
        \end{cases}
    \end{aligned}
\end{equation}
where the first line is a left-moving Minkowski plane wave mode restricted in $v\leq 0$, and the second line recovers a left-moving Milne mode restricted in $0\leq v< v_{0}$ (instead of the standard domain: $-\infty< v<v_{0}$) at $\mathscr{I}_{R}^{-}$.

By analytic continuation across the asymptote $v=v_{0}$, the other pair of the \textit{out} mode $v_{-\omega}^{L}(v>v_{0})$ is obtained (see Fig.~\ref{fig-profile1}):
\begin{equation}\label{rindler-IR-}
    \begin{aligned}
        &\left.v_{-\omega}^{L}(v>v_{0})\right|_{\mathscr{I}_{L}^{+}}
        =
        \left.v_{-\omega}^{L}(v>v_{0})\right|_{\mathscr{I}_{R}^{-}}
        =
        -
        \frac{1}{\sqrt{4\pi\omega}}e^{-\frac{i\omega}{\kappa}\ln(\kappa v-1)}.
    \end{aligned}
\end{equation}

This is nothing but a left-moving Rindler mode ($\omega$ here is the Rindler frequency), which is the well-known entangled partner mode of the Milne mode. It is a complete mode basis and positive frequency with respect to the Rindler time in the standard domain: $v>v_{0}$. According to Eq.~\eqref{bogo-RL-in}, its Bogoliubov coefficients with respect to the left-moving Minkowski plane wave \textit{in} mode $u_{-\omega'}$ are
\begin{equation}\label{bogo-partner-standard}
    \begin{aligned}
        \beta_{-\omega,-\omega'}^{L}
        &=
        -
        \frac{1}{2\pi}\sqrt{\frac{\omega'}{\omega}}
        \int_{v_{0}}^{\infty}dv\, e^{-i\omega'v}
        e^{-\frac{i\omega}{\kappa}\ln(\kappa v-1)}
        \\
        &=
        \frac{ie^{-\frac{i\omega'}{\kappa}}}{2\pi \omega'}\sqrt{\frac{\omega'}{\omega}}e^{-\frac{\pi\omega}{2\kappa}}\big(\frac{\kappa}{\omega'}\big)^{-\frac{i\omega}{\kappa}}
        \Gamma\big(1-\frac{i\omega}{\kappa}\big).
        \\
        \alpha_{-\omega,-\omega'}^{L}
        &=
        -
        \frac{1}{2\pi}\sqrt{\frac{\omega'}{\omega}}
        \int_{v_{0}}^{\infty}dv\, e^{i\omega'v}
        e^{-\frac{i\omega}{\kappa}\ln(\kappa v-1)}
        \\
        &=
        -\frac{ie^{\frac{i\omega'}{\kappa}}}{2\pi \omega'}\sqrt{\frac{\omega'}{\omega}}e^{\frac{\pi\omega}{2\kappa}}\big(\frac{\kappa}{\omega'}\big)^{-\frac{i\omega}{\kappa}}
        \Gamma\big(1-\frac{i\omega}{\kappa}\big),
    \end{aligned}
\end{equation}
which gives the exact spectrum:
\begin{equation}\label{spec-partner-standard}
    \begin{aligned}
        |\beta_{-\omega, -\omega'}^{L}|^{2}
        &=
        \frac{1}{2\pi\kappa \omega'}\frac{1}{e^{\omega/T_{H}}-1},
    \end{aligned}
\end{equation}
which is identical to Eq.~\eqref{spec-hawking-standard}. However, note that the spectrum Eq.~\eqref{spec-hawking-standard} is Bose--Einstein-like in terms of the Minkowski frequency $\omega$, while Eq.~\eqref{spec-partner-standard} is Bose--Einstein-like in terms of the Rindler frequency $\omega$. ($\omega'$ is the Minkowski frequency of the left-moving Minkowski plane wave \textit{in} mode in both cases.)

Finally, by inserting Eqs.~\eqref{bogo-ideal} and \eqref{bogo-partner-standard} into Eq.~\eqref{Q-perfect-asy}, one obtains the standard results \cite{fabbri2005modeling}:
\begin{equation}\label{Q-0}
    \begin{aligned}
        Q^{RR}(\omega_{1},\omega_{2})=Q^{LL}(-\omega_{1},-\omega_{2})=0,
        \quad
        \omega_{1}\neq \omega_{2},
    \end{aligned}
\end{equation}
between different modes. These indicate that the late time right-moving $\{k_{1},k_{2}\}>0$ Minkowski plane wave \textit{out} modes of different frequencies $\omega_{1}\neq\omega_{2}$ are mutually uncorrelated, and the left-moving $\{k_{1},k_{2}\}<0$ Rindler modes of different frequencies $\omega_{1}\neq\omega_{2}$ are also mutually uncorrelated. On the other hand,
\begin{equation}\label{Q-delta}
    \begin{aligned}
        Q^{RL}(k_{1}=\omega_{1},k_{2}=-\omega_{2})
        =
        \frac{[\delta(\omega_{1}-\omega_{2})]^{2}}{4\sinh(\pi\omega_{1}/\kappa)\sinh(\pi\omega_{2}/\kappa)},
    \end{aligned}
\end{equation}
where $\omega_{1}$ is the Minkowski frequency associated with the Minkowski time for the right-moving Minkowski plane wave \textit{out} mode, and $\omega_{2}$ is the Rindler frequency associated with the Rindler time for the left-moving Rindler mode. The appearance of the Dirac delta distribution in $Q^{RL}(k_{1},k_{2})$ indicates that these two particles have a very sharp correlation at $\omega_{1}=\omega_{2}$, i.e., $k_{1}=-k_{2}$. This corresponds to the conventional belief that the semi-classical Hawking radiation (in the sense that the spectrum is given by Eq.~\eqref{bogo-ideal}) is uncorrelated to itself but is only correlated with the partner mode that falls into the horizon of the black hole.

\subsection{Exact Bogoliubov coefficients}

In the previous discussion, the mirror's accelerated motion is extended to the infinite past by hand to obtain the spectrum \eqref{bogo-ideal}, which is valid for all $\{\omega,\omega'\}$. However, we will show in this subsection that the exact evaluation of the Bogoliubov coefficients indicates a deviation from Eq.~\eqref{bogo-ideal} and the existence of nontrivial correlations between the right-moving modes due to the early history of the mirror's motion. By considering the actual trajectory Eq.~\eqref{traj-1}, one obtains the exact Bogoliubov coefficients:
\begin{equation}
    \begin{aligned}
        \beta_{\omega,-\omega'}^{R}
        &=
        \beta_{\omega,-\omega'}^{R,(\text{i})}
        +
        \beta_{\omega,-\omega'}^{R,(\text{ii})},
    \end{aligned}
\end{equation}
where
\begin{equation}\label{beta-perfect-traj-1}
    \begin{aligned}
        \beta_{\omega,-\omega'}^{R,(\text{i})}
        &=
        -
        \frac{1}{2\pi}\sqrt{\frac{\omega'}{\omega}}
        \int_{-\infty}^{0}dT\, e^{-i(\omega+\omega')T}
        \\
        &=
        -
        \frac{1}{2\pi}\sqrt{\frac{\omega'}{\omega}}
        \lim_{\epsilon\rightarrow 0^{+}}
        \frac{i}{\omega+\omega'+i\epsilon}
        =
        -
        \frac{i}{2\pi}\sqrt{\frac{\omega'}{\omega}}
        \frac{1}{\omega+\omega'},
        \\
        \beta_{\omega,-\omega'}^{R,(\text{ii})}
        &=
        -
        \frac{1}{2\pi}\sqrt{\frac{\omega'}{\omega}}
        \int_{0}^{\infty}dT \big(1+Z'(T)\big)
        e^{-i(\omega+\omega')T+i(\omega-\omega')Z(T)}
        \\
        &=
        -
        \frac{ie^{-\frac{i\omega'}{\kappa}}}{2\pi \omega'}\sqrt{\frac{\omega'}{\omega}}
        e^{-\frac{\pi\omega}{2\kappa}}
        \big(\frac{\kappa}{\omega'}\big)^{\frac{i\omega}{\kappa}}
        \big[
        \Gamma
        \big(
        1
        +
        \frac{i\omega}{\kappa}
        \big)
        -
        \Gamma
        \big(
        1
        +
        \frac{i\omega}{\kappa},
        -\frac{i\omega'}{\kappa}
        \big)
        \big],
    \end{aligned}
\end{equation}
and similarly,
\begin{equation}\label{alpha-perfect-traj-1}
    \begin{aligned}
        \alpha_{\omega,-\omega'}^{R,(\text{i})}
        &=
        -
        \frac{1}{2\pi}\sqrt{\frac{\omega'}{\omega}}
        \lim_{\epsilon\rightarrow 0^{+}}
        \frac{i}{\omega-\omega'+i\epsilon},
        \\
        \alpha_{\omega,-\omega'}^{R,(\text{ii})}
        &=
        \frac{ie^{\frac{i\omega'}{\kappa}}}{2\pi \omega'}\sqrt{\frac{\omega'}{\omega}}
        e^{\frac{\pi\omega}{2\kappa}}
        \big(\frac{\kappa}{\omega'}\big)^{\frac{i\omega}{\kappa}}
        \big[
        \Gamma
        \big(
        1
        +
        \frac{i\omega}{\kappa}
        \big)
        -
        \Gamma
        \big(
        1
        +
        \frac{i\omega}{\kappa},
        \frac{i\omega'}{\kappa}
        \big)
        \big].
    \end{aligned}
\end{equation}

This shows that the early stage of static motion not only contributes additional terms $\{\alpha_{\omega,-\omega'}^{R,(\text{i})},\beta_{\omega,-\omega'}^{R,(\text{i})}\}$, but it also introduces upper incomplete Gamma functions to the coefficients $\{\alpha_{\omega,-\omega'}^{R,(\text{ii})},\beta_{\omega,-\omega'}^{R,(\text{ii})}\}$ as a junction effect. Therefore, the frequency spectrum based on Eq.~\eqref{beta-perfect-traj-1} generally deviates from the Bose--Einstein distribution, which only appears in the limits of $\omega'\gg \kappa$ and $\omega'\gg \omega$:
\begin{equation}\label{spec-hawking-perfect}
    \beta_{\omega,-\omega'}^{R}
    \simeq
        -
        \frac{ie^{-\frac{i\omega'}{\kappa}}}{2\pi \omega'}\sqrt{\frac{\omega'}{\omega}}e^{-\frac{\pi\omega}{2\kappa}}\big(\frac{\kappa}{\omega'}\big)^{\frac{i\omega}{\kappa}}
        \Gamma\big(1+\frac{i\omega}{\kappa}\big),
\end{equation}
which reproduces the same expression as Eq.~\eqref{bogo-ideal}, where $\Gamma(x,s)\sim e^{-s}s^{x-1}$ for $|s|\gg 1$ is used so that the upper incomplete Gamma function in $\beta_{\omega,-\omega'}^{R,(\text{ii})}$ cancels with the static contribution $\beta_{\omega,-\omega'}^{R,(\text{i})}$, leaving only the complete gamma function that is relevant for the Bose--Einstein distribution.

On the contrary, if one considers the limit: $\omega'\ll \kappa$, one obtains instead
\begin{equation}\label{beta-perfect-prehawk}
    \beta_{\omega,-\omega'}^{R}
    \simeq
    -
        \frac{i}{2\pi}\sqrt{\frac{\omega'}{\omega}}
        \frac{1}{\omega+\omega'},
\end{equation}
where $\Gamma(x,s)\sim\Gamma(x)-s^{x}/x$ for $|s|\ll 1$ is used, and the complete gamma function is thus eliminated. Since the Bose--Einstein distribution only emerges in certain $\omega'$ regimes, quantities such as the number of particles in a given \textit{out} mode $\omega$, which involve integration over $\omega'$, may deviate from the results evaluated by simply extending the Bose--Einstein distribution over the entire range of $\omega'$. In addition, when investigating correlations among the radiated particles, deviations from the conventional results become relevant.

Finally, for the left-moving Rindler mode, since it is located beyond the asymptote $v=v_{0}$, the motion of the mirror in the prior does not affect the coefficients $\{\alpha_{-\omega,-\omega'}^{L},\beta_{-\omega,-\omega'}^{L}\}$. Therefore, $Q^{LL}(-\omega_{1},-\omega_{2})=0$ remains valid. However, $Q^{RL}(\omega_{1},-\omega_{2})$ and $Q^{RR}(\omega_{1},\omega_{2})$ become more technically involved due to the additional terms in Eqs.~\eqref{beta-perfect-traj-1} and \eqref{alpha-perfect-traj-1}. We shall plot these quantities numerically in the next section.

\subsubsection{Two-mode squeezing of Milne modes}
\label{Single-mode squeezing of Milne modes}

By comparing Eqs.~\eqref{beta-perfect-traj-1} and Eq.~\eqref{alpha-perfect-traj-1} to Eqs.~\eqref{bogo-ideal} and \eqref{bogo-partner-standard}, one immediately expects that the initial motion of the mirror shall affect the correlation between different \textit{out} modes. As we shall see in the next subsection \ref{Additional correlations}, it not only alters the correlation between the right-moving Minkowski plane wave \textit{out} mode $\hat{b}_{\omega}^{R}$ and the left-moving Rindler \textit{out} mode $\hat{b}_{-\omega}^{L}$, but it also induces self-correlation among $\{\hat{b}_{\omega_{1}}^{R},\hat{b}_{\omega_{2}}^{R}\}$ of different frequencies $\omega_{2}\neq \omega_{1}$. Therefore, if one defines \textit{Hawking modes} as the modes that possess a \textit{Bose--Einstein distribution without self-correlations} with respect to the \textit{in} vacuum, then in what sense can one interpret the approximated Bose--Einstein spectrum \eqref{spec-hawking-perfect} as Hawking radiation, and why are the approximations required? 

In the following, we will show that the radiated Minkowski \textit{out} modes $\hat{b}^{R}_{\omega}$ received at $\mathscr{I}_{R}^{+}$ can be interpreted as Milne modes $\hat{a}_{-\Omega'}^{\mathrm{II}}$ that get reflected and squeezed by the moving mirror. Since Milne modes satisfy the definition of Hawking modes defined in the last paragraph, the radiated modes $\hat{b}^{R}_{\omega}$ at $\mathscr{I}_{R}^{+}$ are, in fact, \textit{squeezed} Hawking modes. The squeezing deviates the spectrum from Bose--Einstein and induces additional correlations. If there is no squeezing, the radiated modes $\hat{b}^{R}_{\omega}$ and the reflected Milne modes $\hat{a}_{-\Omega'}^{\mathrm{II}}$ become equivalent.


As mentioned previously, the right-moving particles that possess the Bogoliubov coefficients \eqref{bogo-ideal} are usually regarded as semi-classical Hawking radiation, which has a Bose--Einstein frequency spectrum. For a general mirror trajectory, such a distribution is absent. However, if the trajectory of the mirror possesses a period of acceleration with an asymptote, such as the $T\geq 0$ stage in Eq.~\eqref{traj-1}, then it is still physically reasonable to define Rindler mode and Milne mode located on each side of the asymptote at $\mathscr{I}_{R}^{-}$, and together they form a complete set of mode bases on the Cauchy surface. This allows the \textit{in} vacuum to possess the standard notion of two-mode squeezed state in terms of the Rindler and Milne modes.

The Milne mode is a complete basis in $v<v_{0}$ at $\mathscr{I}_{R}^{-}$, where $v_{0}$ is the asymptote in Eq.~\eqref{raytracing-1-F}. Therefore, one may expand the right-moving Minkowski plane wave \textit{out} mode $\hat{b}_{\omega}^{R}$ in terms of the left-moving Milne mode $\hat{a}_{-\Omega'}^\text{II}$ according to Eq.~\eqref{bogo-RL-I,II}. The occupation number of the right-moving particles observed by the inertial observer at $\mathscr{I}_{R}^{+}$ then becomes
\begin{equation}
    \begin{aligned}
        &\langle 0;in|
        (\hat{b}_{\omega}^{R})^{\dagger}\hat{b}_{\omega}^{R}
        |0;in\rangle
        \\
        &=
        \int_{0}^{\infty}d\Omega_{1}'d\Omega_{2}'
        \Big(
        \alpha_{\omega,-\Omega_{1}'}^\text{II}\bar{\alpha}_{\omega,-\Omega_{2}'}^\text{II}
        \langle 0;in|
        (\hat{a}_{-\Omega_{1}'}^\text{II})^{\dagger}\hat{a}_{-\Omega_{2}'}^\text{II}
        |0;in\rangle
        -
        \alpha_{\omega,-\Omega_{1}'}^\text{II}\bar{\beta}_{\omega,-\Omega_{2}'}^\text{II}
        \langle 0;in|
        (\hat{a}_{-\Omega_{1}'}^\text{II})^{\dagger}(\hat{a}_{-\Omega_{2}'}^\text{II})^{\dagger}
        |0;in\rangle
        \\
        &\quad\quad\quad\quad\quad\quad\quad
        -
        \beta_{\omega,-\Omega_{1}'}^\text{II}\bar{\alpha}_{\omega,-\Omega_{2}'}^\text{II}
        \langle 0;in|
        \hat{a}_{-\Omega_{1}'}^\text{II}\hat{a}_{-\Omega_{2}'}^\text{II}
        |0;in\rangle
        +
        \beta_{\omega,-\Omega_{1}'}^\text{II}\bar{\beta}_{\omega,-\Omega_{2}'}^\text{II}
        \langle 0;in|
        \hat{a}_{-\Omega_{1}'}^\text{II}(\hat{a}_{-\Omega_{2}'}^\text{II})^{\dagger}
        |0;in\rangle
        \Big).
    \end{aligned}
\end{equation}

\begin{figure}[h]
    \centering
    \includegraphics[width=0.5\linewidth]{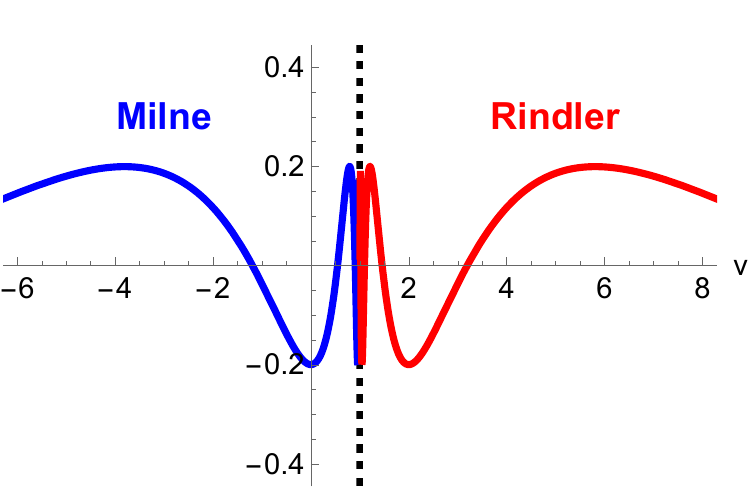}  
    \caption{Spacetime profiles of the Milne mode and the Rindler mode at the past null infinity by taking the real part of Eq.~\eqref{rindlermile-mode-past}. The vertical black dashed line indicates the asymptote $v_{0}=1/\kappa$. We set $\kappa=1,\Omega'=2$. }
    \label{fig-profile2}
\end{figure}

To evaluate the above quantities, let us expand the Rindler/Milne modes by the Minkowski \textit{in} modes by
\begin{equation}\label{in-rindlermile-mode}
    \begin{aligned}
        u_{-\Omega'}^\text{I}
        &=
        \int_{0}^{\infty}d\omega'
        \big(
        \tilde{\alpha}_{-\Omega',-\omega'}^\text{I}
        u_{-\omega' }
        +
        \tilde{\beta}_{-\Omega',-\omega'}^\text{I}
        \bar{u}_{-\omega' }
        \big),
        \\
        u_{-\Omega'}^\text{II}
        &=
        \int_{0}^{\infty}d\omega'
        \big(
        \tilde{\alpha}_{-\Omega',-\omega'}^\text{II}
        u_{-\omega' }
        +
        \tilde{\beta}_{-\Omega',-\omega'}^\text{II}
        \bar{u}_{-\omega' }
        \big),
        \\
        \hat{a}_{-\Omega'}^\text{I}
        &=
        \int_{0}^{\infty}d\omega' 
        \big(
        \bar{\tilde{\alpha}}_{-\Omega',-\omega'}^\text{I}
        \hat{a}_{-\omega' }
        -
        \bar{\tilde{\beta}}_{-\Omega',-\omega'}^\text{I}
        \hat{a}_{-\omega' }^{\dagger}
        \big),
        \\
        \hat{a}_{-\Omega'}^\text{II}
        &=
        \int_{0}^{\infty}d\omega' 
        \big(
        \bar{\tilde{\alpha}}_{-\Omega',-\omega'}^\text{II}
        \hat{a}_{-\omega' }
        -
        \bar{\tilde{\beta}}_{-\Omega',-\omega'}^\text{II}
        \hat{a}_{-\omega' }^{\dagger}
        \big),
    \end{aligned}
\end{equation}
where (see Fig.~\ref{fig-profile2})
\begin{equation}\label{rindlermile-mode-past}
    \begin{aligned}
        \left.u_{-\Omega'}^\text{I}\right|_{\mathscr{I}_{R}^{-}}
        &=
        -
        \frac{\Theta(v-v_{0})}{\sqrt{4\pi\Omega'}}
        e^{-\frac{i\Omega'}{\kappa}\ln(\kappa v-1)},
        \\
        \left.u_{-\Omega'}^\text{II}\right|_{\mathscr{I}_{R}^{-}}
        &=
        -
        \frac{\Theta(v_{0}-v)}{\sqrt{4\pi\Omega'}}
        e^{\frac{i\Omega'}{\kappa}\ln(1-\kappa v)},
    \end{aligned}
\end{equation}
and
\begin{equation}\label{bogo-tilde}
    \begin{aligned}
        \tilde{\alpha}_{-\Omega',-\omega'}^\text{I}
        &=
        -
        \frac{1}{2\pi}\sqrt{\frac{\omega'}{\Omega'}}
        \int_{v_{0}}^{\infty}dv\, e^{i\omega'v}
        e^{-\frac{i\Omega'}{\kappa}\ln(\kappa v-1)}
        \\
        &=
        -\frac{ie^{\frac{i\omega'}{\kappa}}}{2\pi \omega'}\sqrt{\frac{\omega'}{\Omega'}}e^{\frac{\pi\Omega'}{2\kappa}}\big(\frac{\kappa}{\omega'}\big)^{-\frac{i\Omega'}{\kappa}}
        \Gamma\big(1-\frac{i\Omega'}{\kappa}\big),
        \\
        \tilde{\beta}_{-\Omega',-\omega'}^\text{I}
        &=
        -
        \frac{1}{2\pi}\sqrt{\frac{\omega'}{\Omega'}}
        \int_{v_{0}}^{\infty}dv\, e^{-i\omega'v}
        e^{-\frac{i\Omega'}{\kappa}\ln(\kappa v-1)}
        \\
        &=
        \frac{ie^{-\frac{i\omega'}{\kappa}}}{2\pi \omega'}\sqrt{\frac{\omega'}{\Omega'}}e^{-\frac{\pi\Omega'}{2\kappa}}\big(\frac{\kappa}{\omega'}\big)^{-\frac{i\Omega'}{\kappa}}
        \Gamma\big(1-\frac{i\Omega'}{\kappa}\big),
        \\
        \tilde{\alpha}_{-\Omega',-\omega'}^\text{II}
        &=
        -
        \frac{1}{2\pi}\sqrt{\frac{\omega'}{\Omega'}}
        \int_{-\infty}^{v_{0}}dv\, e^{i\omega'v}
        e^{\frac{i\Omega'}{\kappa}\ln(1-\kappa v)}
        \\
        &=
        \frac{ie^{\frac{i\omega'}{\kappa}}}{2\pi \omega'}\sqrt{\frac{\omega'}{\Omega'}}e^{\frac{\pi\Omega'}{2\kappa}}\big(\frac{\kappa}{\omega'}\big)^{\frac{i\Omega'}{\kappa}}
        \Gamma\big(1+\frac{i\Omega'}{\kappa}\big),
        \\
        \tilde{\beta}_{-\Omega',-\omega'}^\text{II}
        &=
        -
        \frac{1}{2\pi}\sqrt{\frac{\omega'}{\Omega'}}
        \int_{-\infty}^{v_{0}}dv\, e^{-i\omega'v}
        e^{\frac{i\Omega'}{\kappa}\ln(1-\kappa v)}
        \\
        &=
        -
        \frac{ie^{-\frac{i\omega'}{\kappa}}}{2\pi \omega'}\sqrt{\frac{\omega'}{\Omega'}}e^{-\frac{\pi\Omega'}{2\kappa}}\big(\frac{\kappa}{\omega'}\big)^{\frac{i\Omega'}{\kappa}}
        \Gamma\big(1+\frac{i\Omega'}{\kappa}\big).
    \end{aligned}
\end{equation}

\begin{figure}[h]
    \centering
    \includegraphics[width=0.5\linewidth]{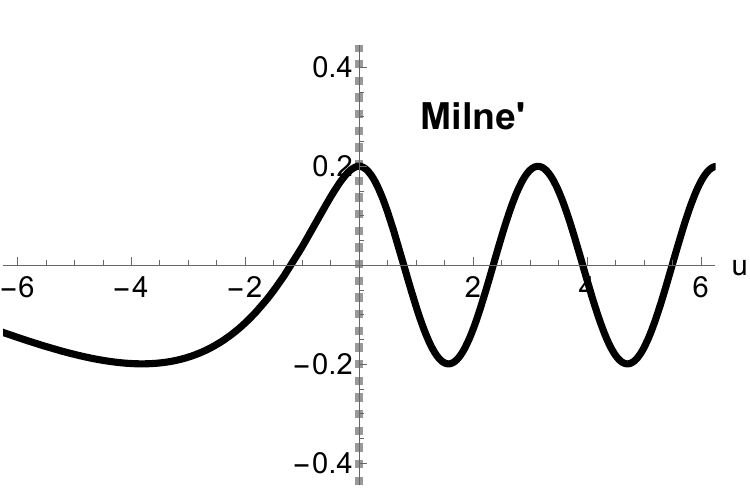}  
    \caption{Spacetime profile of the Milne mode after reflection at the future null infinity by taking the real part of Eq.~\eqref{milne-at-future}. The vertical gray dashed line indicates the transition moment of the mirror's motion. We set $\kappa=1,\Omega'=2$. }
    \label{fig-profile3}
\end{figure}

With these relations and using Eq.~\eqref{bogo-uI-II,vRL} and $\hat a_{-\omega'}|0;in\rangle=0$ for $\omega'>0$,  one obtains (using Appendix~\ref{A1})
\begin{equation}\label{spectrum-squeezed-1}
    \begin{aligned}
        \langle 0;in|
        (\hat{b}_{\omega}^{R})^{\dagger}\hat{b}_{\omega}^{R}
        |0;in\rangle
        =
        \int_{0}^{\infty}d\Omega'
        \frac{|\alpha_{\omega,-\Omega'}^\text{II}|^{2} }{e^{\frac{2\pi\Omega'}{\kappa}}-1}
        \Biggl(
        1
        +
        e^{\frac{2\pi\Omega'}{\kappa}}
        \frac{|\beta_{\omega,-\Omega'}^\text{II}|^{2}}{|\alpha_{\omega,-\Omega'}^\text{II}|^{2}}
        \Biggr),
    \end{aligned}
\end{equation}
which shows that the frequency $\omega$-spectrum of the right-moving radiation defined by the Minkowski plane wave \textit{out} mode $\hat{b}_{\omega}^{R}$ deviates from the Bose--Einstein distribution due to the non-trivial two-mode squeezing of the Milne modes (Hawking modes) originated from $\{\alpha_{\omega,-\Omega'}^\text{II},\beta_{\omega,-\Omega'}^\text{II}\}$, which depends on the entire history $(-\infty<v\leq v_{0})$ of the mirror's motion. Notably, the above expression elegantly separates the pure Bose--Einstein behavior of the Milne mode with respect to $|0;in\rangle$ from the mirror-trajectory's history-dependent squeezing. Explicitly, according to Eq.~\eqref{bogo-uI-II,vRL}, 
\begin{equation}
    \begin{aligned}
        u_{-\Omega'}^\text{II}
        &=
        \int_{0}^{\infty}d\omega
        \big(
        \bar{\alpha}_{\omega,-\Omega'}^\text{II}
        v_{\omega}^{R}
        -
        \beta_{\omega,-\Omega'}^\text{II}
        \bar{v}_{\omega}^{R}
        \big),
    \end{aligned}
\end{equation}
and evaluating the Bogoliubov coefficients at $\mathscr{I}_{R}^{+}$ using (see Fig.~\ref{fig-profile3})
\begin{equation}\label{milne-at-future}
    \begin{aligned}
        \left.u_{-\Omega'}^\text{II}\right|_{\mathscr{I}_{R}^{+}}
        &=
        \begin{cases}
            \frac{1}{\sqrt{4\pi\Omega'}}e^{\frac{i\Omega'}{\kappa}\ln(1-\kappa u)},
            \quad
            -\infty< u\leq 0
            \\
            \frac{1}{\sqrt{4\pi\Omega'}}e^{-i\Omega' u},
            \quad
            0\leq u<\infty,
        \end{cases}
    \end{aligned}
\end{equation}
one obtains
\begin{equation}
    \begin{aligned}
        \bar{\alpha}_{\omega,-\Omega'}^\text{II}
        &=
        \bar{\alpha}_{\omega,-\Omega'}^\text{II,(i)}
        +
        \bar{\alpha}_{\omega,-\Omega'}^\text{II,(ii)},
        \\
        \beta_{\omega,-\Omega'}^\text{II}
        &=
        \beta_{\omega,-\Omega'}^\text{II,(i)}
        +
        \beta_{\omega,-\Omega'}^\text{II,(ii)},
    \end{aligned}
\end{equation}
where
\begin{equation}
    \begin{aligned}
        \bar{\alpha}_{\omega,-\Omega'}^\text{II,(i)}
        &=
        \frac{1}{2\pi}
        \sqrt{\frac{\omega}{\Omega'}}
        \int_{-\infty}^{0} du\,
        e^{i\omega u}
        e^{\frac{i\Omega'}{\kappa}\ln(1-\kappa u)},
        \\
        \bar{\alpha}_{\omega,-\Omega'}^\text{II,(ii)}
        &=
        \frac{1}{2\pi}
        \sqrt{\frac{\omega}{\Omega'}}
        \int_{0}^{\infty} du\,
        e^{i\omega u}
        e^{-i\Omega' u},
        \\
        \beta_{\omega,-\Omega'}^\text{II,(i)}
        &=
        -
        \frac{1}{2\pi}
        \sqrt{\frac{\omega}{\Omega'}}
        \int_{-\infty}^{0} du\,
        e^{-i\omega u}
        e^{\frac{i\Omega'}{\kappa}\ln(1-\kappa u)},
        \\
        \beta_{\omega,-\Omega'}^\text{II,(ii)}
        &=
        -
        \frac{1}{2\pi}
        \sqrt{\frac{\omega}{\Omega'}}
        \int_{0}^{\infty} du\,
        e^{-i\omega u}
        e^{-i\Omega' u},
    \end{aligned}
\end{equation}
    where the mirror's history splits the integration into two regimes: $u\in(-\infty,0]$ and $u\in[0,\infty)$. This clearly distorts the Bose--Einstein distribution in Eq. \eqref{spectrum-squeezed-1}.

In the case that the accelerated motion in Eq.~\eqref{traj-1} is extended to the infinite past, there is no squeezing since $\{\alpha_{\omega,-\Omega'}^\text{II},\beta_{\omega,-\Omega'}^\text{II}\}=\{\delta(\omega-\Omega'),0\}$ and the ideal Bose--Einstein distribution is recovered:
\begin{equation}
    \begin{aligned}
        &\langle 0;in|
        (\hat{b}_{\omega}^{R})^{\dagger}\hat{b}_{\omega}^{R}
        |0;in\rangle
        =
        \frac{\delta(0)}{e^{2\pi\omega/\kappa}-1},
    \end{aligned}
\end{equation}
where the Dirac delta, $\delta(0)=\delta(\omega-\omega)=\Delta T/2\pi\rightarrow \infty$ indicates eternal acceleration, $\Delta T\rightarrow\infty$.

\subsubsection{Additional correlations}
\label{Additional correlations}

Previously, we briefly mentioned the correlation between various modes in the moving mirror model with an asymptotically null motion in terms of the Bogoliubov coefficients $\{\alpha_{\omega,-\omega'}^{R},\beta_{\omega,-\omega'}^{R},\alpha_{-\omega,-\omega'}^{L},\beta_{-\omega,-\omega'}^{L}\}$ that relate the right-moving Minkowski plane wave \textit{out} mode and the left-moving Minkowski plane wave \textit{in} mode, and the left-moving Rindler \textit{out} mode and the left-moving Minkowski plane wave \textit{in} mode. Although it is possible to evaluate the Bogoliubov coefficients analytically, it is not transparent to gain a physical understanding of the correlation in terms of the Hawking mode and its partner mode. However, now that we are aware of the two-mode squeezing of Milne modes, one may investigate how the squeezing affects the correlations.

According to Eq.~\eqref{Q-perfect-asy}, the correlations are essentially determined by $\{\alpha_{\omega,-\omega'}^{R},\beta_{\omega,-\omega'}^{R},\alpha_{-\omega,-\omega'}^{L},\beta_{-\omega,-\omega'}^{L}\}$. Thus, our strategy is to express these Bogoliubov coefficients in terms of those in Eq.~\eqref{bogo-tilde}:
\begin{equation}\label{bogo-additional1}
    \begin{aligned}
        \alpha_{\omega,-\omega'}^{R}
        &=
        -
        \frac{1}{2\pi}
        \sqrt{\frac{\omega'}{\omega}}
        \int_{-\infty}^{v_{0}}dv\,
        e^{i\omega' v}e^{-i\omega F(v)}
        \\
        &=
        \int_{0}^{\infty}d\Omega'
        \big(
        \alpha_{\omega,-\Omega'}^\text{II}
        \tilde{\alpha}_{-\Omega',-\omega'}^\text{II}
        +
        \beta_{\omega,-\Omega'}^\text{II}
        \bar{\tilde{\beta}}_{-\Omega',-\omega'}^\text{II}
        \big),
        \\
        \beta_{\omega,-\omega'}^{R}
        &=
        -
        \frac{1}{2\pi}
        \sqrt{\frac{\omega'}{\omega}}
        \int_{-\infty}^{v_{0}}dv\,
        e^{-i\omega' v}e^{-i\omega F(v)}
        \\
        &=
        \int_{0}^{\infty}d\Omega'
        \big(
        \alpha_{\omega,-\Omega'}^\text{II}
        \tilde{\beta}_{-\Omega',-\omega'}^\text{II}
        +
        \beta_{\omega,-\Omega'}^\text{II}
        \bar{\tilde{\alpha}}_{-\Omega',-\omega'}^\text{II}
        \big),
        \\
        \alpha_{-\omega,-\omega'}^{L}
        &=
        \tilde{\alpha}_{-\omega,-\omega'}^\text{I}
        =
        -\bar{\tilde{\beta}}_{-\omega,-\omega'}^\text{II}e^{\frac{\pi\omega}{\kappa}},
        \\
        \beta_{-\omega,-\omega'}^{L}
        &=
        \tilde{\beta}_{-\omega,-\omega'}^\text{I}
        =
        -\bar{\tilde{\alpha}}_{-\omega,-\omega'}^\text{II}e^{-\frac{\pi\omega}{\kappa}}.
    \end{aligned}
\end{equation}

With these expressions, one obtains the following expressions (using Appendix~\ref{A2}):
    \begin{align}
        &Q^{RR}(\omega_{1},\omega_{2}) \notag
        \\
        &=
        \int_{0}^{\infty}d\Omega_{1}'
        \frac{e^{-\frac{\pi\Omega_{1}'}{\kappa}}\alpha_{\omega_{1},-\Omega_{1}'}^\text{II}\bar{\alpha}_{\omega_{2},-\Omega_{1}'}^\text{II}+e^{\frac{\pi\Omega_{1}'}{\kappa}}\beta_{\omega_{1},-\Omega_{1}'}^\text{II}\bar{\beta}_{\omega_{2},-\Omega_{1}'}^\text{II} }{2\sinh(\pi\Omega_{1}'/\kappa)}
        \int_{0}^{\infty}d\Omega_{2}'
        \frac{e^{\frac{\pi\Omega_{2}'}{\kappa}}\bar{\alpha}_{\omega_{1},-\Omega_{2}'}^\text{II}\alpha_{\omega_{2},-\Omega_{2}'}^\text{II}+e^{-\frac{\pi\Omega_{2}'}{\kappa}}\bar{\beta}_{\omega_{1},-\Omega_{2}'}^\text{II}\beta_{\omega_{2},-\Omega_{2}'}^\text{II} }{2\sinh(\pi\Omega_{2}'/\kappa)} \notag
        \\
        &+
        \int_{0}^{\infty}d\Omega_{1}'
        \frac{e^{-\frac{\pi\Omega_{1}'}{\kappa}}\alpha_{\omega_{1},-\Omega_{1}'}^\text{II}\beta_{\omega_{2},-\Omega_{1}'}^\text{II}+e^{\frac{\pi\Omega_{1}'}{\kappa}}\beta_{\omega_{1},-\Omega_{1}'}^\text{II}\alpha_{\omega_{2},-\Omega_{1}'}^\text{II} }{2\sinh(\pi\Omega_{1}'/\kappa)}
        \int_{0}^{\infty}d\Omega_{2}'
        \frac{e^{\frac{\pi\Omega_{2}'}{\kappa}}\bar{\alpha}_{\omega_{1},-\Omega_{2}'}^\text{II}\bar{\beta}_{\omega_{2},-\Omega_{2}'}^\text{II}+e^{-\frac{\pi\Omega_{2}'}{\kappa}}\bar{\beta}_{\omega_{1},-\Omega_{2}'}^\text{II}\bar{\alpha}_{\omega_{2},-\Omega_{2}'}^\text{II} }{2\sinh(\pi\Omega_{2}'/\kappa)}, \label{correlation-squeezed-1}
        \\
        &Q^{RL}(\omega_{1},-\omega_{2})
        =
        \frac{\Big(
        |\beta_{\omega_{1},-\omega_{2}}^\text{II}|^{2}
        +
        |\alpha_{\omega_{1},-\omega_{2}}^\text{II}|^{2}
        \Big)}{4\sinh^{2}(\pi\omega_{2}/\kappa)}, \label{correlation-squeezed-2}
    \end{align}
which clearly separate the contribution from the Milne mode itself (pure $\Omega'$ terms) from the two-mode squeezing effect determined by the history of the mirror's trajectory. From the above expressions, one also observes that Eqs.~\eqref{Q-0} and \eqref{Q-delta} are recovered in the absence of two-mode squeezing. Otherwise, the weakening in the correlation $Q^{RL}$ due to the squeezing, i.e., the numerator is no longer $[\delta(\omega_1-\omega_2)]^2$, is accompanied by an increase in the correlation $Q^{RR}$ between the right-moving particles.

\section{Formation and evaporation of black holes}
\label{Formation and evaporation of black holes}

\subsection{Perspective of inertial observers at $\mathscr{I}_{R}^{+}$}

In the scenario where the formed black hole undergoes subsequent evaporation to lose its mass, one may consider a mirror that initially undergoes a period of acceleration, i.e., black hole formation, and eventually returns to rest, i.e., complete evaporation. For analytic simplicity, we consider the following trajectory (see Fig.~\ref{fig-trajectory}):
\begin{equation}\label{traj-2}
    \begin{aligned}
        Z(T)
        =
        \begin{cases}
            0,
            \quad
            T\leq 0
            \\
            -T+\frac{1}{\kappa}-\frac{1}{\kappa}W\big(e^{1-2\kappa T}\big),
            \quad
            0\leq T \leq T_{*}
            \\
            -\frac{1}{2}(f_{*}-v_{*}),
            \quad
            T_{*}\leq T,
        \end{cases}
    \end{aligned}
\end{equation}
where $W(x)$ is the Lambert W function, $\{\kappa,v_{*}\}$ are parameters of the trajectory related to the acceleration and the moment of deceleration, respectively, and 
\begin{equation}
    \begin{aligned}
        T_{*}
        =
        \frac{1}{2}(v_{*}+f_{*}),
        \quad
        f_{*}
        =
        -\frac{1}{\kappa}\ln(1-\kappa v_{*}),
    \end{aligned}
\end{equation}
where $T=T_{*}$ is the moment when the mirror returns to rest.

In this case, the corresponding Bogoliubov coefficients for a moving mirror with perfect reflectivity, according to the inertial observer at $\mathscr{I}_{R}^{+}$, are
\begin{equation}
    \begin{aligned}
        \beta_{\omega,-\omega'}^{R}
        &=
        \beta_{\omega,-\omega'}^{R,(\text{i})}
        +
        \beta_{\omega,-\omega'}^{R,(\text{iii})}
        +
        \beta_{\omega,-\omega'}^{R,(\text{ii})},
    \end{aligned}
\end{equation}
where
\begin{equation}\label{beta-2}
    \begin{aligned}
        \beta_{\omega,-\omega'}^{R,(\text{i})}
        &=
        -
        \frac{1}{2\pi}\sqrt{\frac{\omega'}{\omega}}
        \int_{-\infty}^{0}dT e^{-i(\omega+\omega')T}
        =
        -
        \frac{i}{2\pi}\sqrt{\frac{\omega'}{\omega}}
        \frac{1}{\omega+\omega'},
        \\
        \beta_{\omega,-\omega'}^{R,(\text{ii})}
        &=
        -
        \frac{1}{2\pi}\sqrt{\frac{\omega'}{\omega}}
        \int_{0}^{T_{*}}dT \big(1+Z'(T)\big)
        e^{-i(\omega+\omega')T+i(\omega-\omega')Z(T)}
        \\
        &=
        -
        \frac{ie^{-\frac{i\omega'}{\kappa}}}{2\pi \omega'}\sqrt{\frac{\omega'}{\omega}}
        e^{-\frac{\pi\omega}{2\kappa}}
        \big(\frac{\kappa}{\omega'}\big)^{\frac{i\omega}{\kappa}}
        \big[
        \Gamma
        \big(
        1
        +
        \frac{i\omega}{\kappa},
        -\frac{i\omega' s_{*}}{\kappa}
        \big)
        -
        \Gamma
        \big(
        1
        +
        \frac{i\omega}{\kappa},
        -\frac{i\omega'}{\kappa}
        \big)
        \big],
        \\
        \beta_{\omega,-\omega'}^{R,(\text{iii})}
        &=
        -
        \frac{1}{2\pi}\sqrt{\frac{\omega'}{\omega}}
        \int_{T_{*}}^{\infty}dT e^{-i(\omega+\omega')T+i(\omega-\omega')Z_{*}}
        =
        \frac{i}{2\pi}\sqrt{\frac{\omega'}{\omega}}\frac{e^{-i(\omega+\omega')T_{*}+i(\omega-\omega')Z_{*}}}{\omega+\omega'},
    \end{aligned}
\end{equation}
where $s_{*}:=W(e^{1-2\kappa T_{*}})$, $Z_{*}:=Z(T_{*})$, and similarly,
\begin{equation}\label{alpha-2}
    \begin{aligned}
        \alpha_{\omega,-\omega'}^{R,(\text{i})}
        &=
        -
        \frac{i}{2\pi}\sqrt{\frac{\omega'}{\omega}}
        \frac{1}{\omega-\omega'+i0^{+}},
        \\
        \alpha_{\omega,-\omega'}^{R,(\text{ii})}
        &=
        \frac{ie^{\frac{i\omega'}{\kappa}}}{2\pi \omega'}\sqrt{\frac{\omega'}{\omega}}
        e^{\frac{\pi\omega}{2\kappa}}
        \big(\frac{\kappa}{\omega'}\big)^{\frac{i\omega}{\kappa}}
        \big[
        \Gamma
        \big(
        1
        +
        \frac{i\omega}{\kappa},
        \frac{i\omega' s_{*}}{\kappa}
        \big)
        -
        \Gamma
        \big(
        1
        +
        \frac{i\omega}{\kappa},
        \frac{i\omega'}{\kappa}
        \big)
        \big],
        \\
        \alpha_{\omega,-\omega'}^{R,(\text{iii})}
        &=
        \frac{i}{2\pi}\sqrt{\frac{\omega'}{\omega}}\frac{e^{-i(\omega-\omega')T_{*}+i(\omega+\omega')Z_{*}}}{\omega-\omega'+i0^{+}}.
    \end{aligned}
\end{equation}

With the inclusion of the additional static motion after $T\geq T_{*}$, the Bogoliubov coefficients not only acquire the additional pieces $\{\alpha_{\omega,-\omega'}^{R,(\text{iii})},\beta_{\omega,-\omega'}^{R,(\text{iii})}\}$, but they also alter the supposed complete gamma functions into incomplete ones due to the introduction of the parameter $s_{*}$. If one takes similar limits as in the previous section: $\omega'\gg \kappa$, $\omega'\gg \omega$, and \(\omega|T_*-Z_*|\ll1\), one obtains
\begin{equation}
    \begin{aligned}
        &\beta_{\omega,-\omega'}^{R}
        \simeq
        \frac{i}{2\pi}\sqrt{\frac{\omega'}{\omega}}\frac{e^{-i\omega' T_{*}-i\omega' Z_{*}}}{\omega'}
        -
        \frac{ie^{-\frac{i\omega'}{\kappa}}}{2\pi \omega'}\sqrt{\frac{\omega'}{\omega}}
        e^{-\frac{\pi\omega}{2\kappa}}
        \big(\frac{\kappa}{\omega'}\big)^{\frac{i\omega}{\kappa}}
        \Gamma
        \big(
        1
        +
        \frac{i\omega}{\kappa},
        -\frac{i\omega' s_{*}}{\kappa}
        \big),
    \end{aligned}
\end{equation}
instead of the approximated Bose--Einstein spectrum \eqref{spec-hawking-perfect}. Therefore, for a given stopping trajectory $(s_{*}\neq 0)$, the Bose--Einstein distribution can no longer be recovered by imposing certain limits on the frequencies $\{\omega,\omega'\}$. For example, if one further takes the limit: $\omega's_{*} \ll \kappa$ for a given $s_{*}\neq 0$, one obtains
\begin{equation}\label{bogo-deviate}
    \begin{aligned}
        \beta_{\omega,-\omega'}^{R}
        &\simeq
        \frac{i}{2\pi}\sqrt{\frac{\omega'}{\omega}}\frac{e^{-i\omega' T_{*}-i\omega' Z_{*}}}{\omega'}
        +
        \frac{e^{-\frac{i\omega'}{\kappa}}}{2\pi}
        \sqrt{\frac{\omega'}{\omega}}
        \frac{s_{*}^{1+\frac{i\omega}{\kappa}}}{\kappa+i\omega}
        -
        \frac{ie^{-\frac{i\omega'}{\kappa}}}{2\pi \omega'}\sqrt{\frac{\omega'}{\omega}}
        e^{-\frac{\pi\omega}{2\kappa}}
        \big(\frac{\kappa}{\omega'}\big)^{\frac{i\omega}{\kappa}}
        \Gamma
        \big(
        1
        +
        \frac{i\omega}{\kappa}
        \big)
        \\
        &\simeq
        \frac{i}{2\pi}\sqrt{\frac{\omega'}{\omega}}\frac{e^{-i\omega' T_{*}-i\omega' Z_{*}}}{\omega'}
        -
        \frac{ie^{-\frac{i\omega'}{\kappa}}}{2\pi \omega'}\sqrt{\frac{\omega'}{\omega}}
        e^{-\frac{\pi\omega}{2\kappa}}
        \big(\frac{\kappa}{\omega'}\big)^{\frac{i\omega}{\kappa}}
        \Gamma
        \big(
        1
        +
        \frac{i\omega}{\kappa}
        \big),
    \end{aligned}
\end{equation}
where the remaining first term in the last line spoils the Bose--Einstein behavior. The first term vanishes only when $s_{*}\rightarrow 0$ and $\omega' \rightarrow \omega'-i0^{+}$, but this simply corresponds to the case in which the mirror accelerates forever.

\subsection{Rindler-Milne modes in disguise}

In the previous section, we employed the asymptotically null trajectory, Eq.~\eqref{traj-1}, to demonstrate that the scalar particles radiated to the mirror's right-hand side are composed of squeezed Milne modes according to an inertial \textit{out} observer at $\mathscr{I}_{R}^{+}$. If the mirror returns to rest after the acceleration, there is no longer a true asymptote, and all the Minkowski plane wave \textit{in} modes will be reflected by the mirror. In this case, both the Milne mode and the Rindler mode will be accessible at $\mathscr{I}_{R}^{+}$. According to our previous discussion on the squeezing of Milne modes, it is then natural to expect that the Rindler modes will also arrive at $\mathscr{I}_{R}^{+}$ in the form of a squeezed state. As we will demonstrate below, this is indeed true. In fact, in addition to the individual two-mode squeezing of the Rindler/Milne modes, there will also be mutual two-mode squeezing between them. Therefore, what the inertial observer at $\mathscr{I}_{R}^{+}$ observes is a mixture of \textit{multiple squeezings of the Rindler/Milne modes.}

\subsubsection{Radiation spectrum observed at $\mathscr{I}_{R}^{+}$}

To study how this mixture manifests in the frequency spectrum of the inertial \textit{out} particles, let us work in the lightcone coordinates for convenience. The ray-tracing function for Eq.~\eqref{traj-2} (aymptotically timelike mirror) is (see Fig. \ref{fig-combined1})
\begin{equation}\label{raytracing-2-F}
    \begin{aligned}
        F(v)
        &=
        \begin{cases}
            v,
            \quad
            v\leq 0
            \\
            -\frac{1}{\kappa}\ln(1-\kappa v),
            \quad
            0\leq v \leq v_{*}<v_{0}
            \\
            f_{*}+(v-v_{*}),
            \quad
            v_{*}\leq v.
        \end{cases}
    \end{aligned}
\end{equation}

In this case, $v_{0}=1/\kappa$ is no longer a real asymptote of the mirror's trajectory since the mirror returns to rest at $v_{*}$ before reaching $v_{0}$. However, it is still physically reasonable to partition the \textit{in} region $\mathscr{I}_{R}^{-}$ into two subregions by the supposed asymptote $v_{0}$. In each subregion, a field operator can be expanded either by the Rindler mode $\hat{a}_{-\Omega'}^\text{I}$ in $v_{0}<v<\infty$ or the Milne mode $\hat{a}_{-\Omega'}^\text{II}$ in $-\infty<v<v_{0}$, since they are complete bases in each subregion. However, note that in the previous trajectory, Eq.~\eqref{traj-1}, we interpreted the Milne mode as the Hawking mode because its spectrum is exactly Bose--Einstein and self-uncorrelated (\textit{c.f.} its partner mode, i.e., the Rindler mode, which also has a Bose--Einstein spectrum). In that case, the Milne mode is only squeezed by the mirror's motion before and during the acceleration according to the inertial \textit{out} observer at $\mathscr{I}_{R}^{+}$. However, in Eq.~\eqref{raytracing-2-F}, the \textit{out} vacuum defined by the inertial \textit{out} observer at $\mathscr{I}_{R}^{+}$ is no longer solely a two-mode squeezed state of the Milne modes. Instead, it involves the squeezing of both Milne modes and Rindler modes. To see this, we again employ the mode expansion according to Eqs.~\eqref{in-rindlermile-mode} and \eqref{rindlermile-mode-past}.

\begin{figure}[h]
    \centering
    \includegraphics[width=0.5\linewidth]{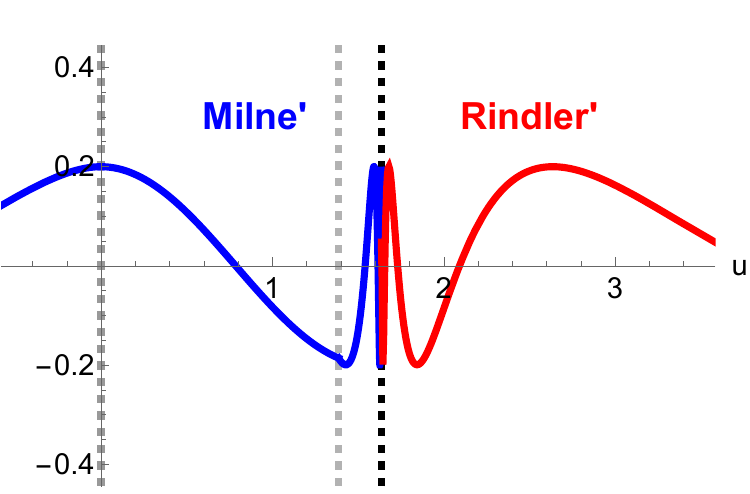}  
    \caption{Spacetime profiles of the Milne mode and the Rindler mode after reflection at the future null infinity by taking the real part of Eq.~\eqref{milne,rindler,after reflection}. The vertical black dashed line indicates $u_{0}$, the reflection of the asymptote $v_{0}=1/\kappa$, and the vertical gray dashed lines are the transition moments $(u=\{0,u_{*}\})$ of the mirror's motion. We set $\kappa=1, \Omega'=2, v_{*}=0.75$. }
    \label{fig-profile4}
\end{figure}

Upon reflection of the left-moving Rindler/Milne modes at the mirror $v=P(u)=F^{-1}(u)$:
\begin{equation}\label{raytracing-2-P}
    \begin{aligned}
        P(u)
        &=
        \begin{cases}
            u,
            \quad
            u\leq 0
            \\
            \frac{1}{\kappa}-\frac{1}{\kappa}e^{-\kappa u},
            \quad
            0\leq u \leq u_{*}<u_{0}
            \\
            u-(f_{*}-v_{*}),
            \quad
            u_{*}\leq u,
        \end{cases}
    \end{aligned}
\end{equation}
where $u_{*}:=f_{*},u_{0}:=F(v_{0})=f_{*}+(v_{0}-v_{*})$, the reflected Rindler/Milne modes have the expressions (see Fig.~\ref{fig-profile4}):
\begin{equation}\label{milne,rindler,after reflection}
    \begin{aligned}
        \left.u_{-\Omega'}^\text{I}\right|_{\mathscr{I}_{R}^{+}}
        &=
        \frac{1}{\sqrt{4\pi\Omega'}}\Theta(u-u_{0})e^{-\frac{i\Omega'}{\kappa}\ln(\kappa(u-u_{0}))},
        \\
        \left.u_{-\Omega'}^\text{II}\right|_{\mathscr{I}_{R}^{+}}
        &=
        \begin{cases}
            \frac{1}{\sqrt{4\pi\Omega'}}e^{\frac{i\Omega'}{\kappa}\ln(1-\kappa u)},
            \quad
            u\leq 0
            \\
            \frac{1}{\sqrt{4\pi\Omega'}}e^{-i\Omega' u},
            \quad
            0\leq u \leq u_{*}
            \\
            \frac{1}{\sqrt{4\pi\Omega'}}e^{\frac{i\Omega'}{\kappa}\ln(\kappa(u_{0}-u))},
            \quad
            u_{*}\leq u<u_{0}.
        \end{cases}
    \end{aligned}
\end{equation}

One may expand the reflected Rindler/Milne modes by the right-moving Minkowski plane wave \textit{out} mode according to the inertial observer at $\mathscr{I}_{R}^{+}$ by the following Bogoliubov transformations:
\begin{equation}
    \begin{aligned}
        u_{-\Omega'}^\text{I}
        &=
        \int_{0}^{\infty}d\omega
        \big(
        \gamma_{-\Omega',\omega}^{\text{I}}v_{\omega}^{R}
        +
        \delta_{-\Omega',\omega}^{\text{I}}\bar{v}_{\omega}^{R}
        \big),
        \\
        u_{-\Omega'}^\text{II}
        &=
        \int_{0}^{\infty}d\omega
        \big(
        \gamma_{-\Omega',\omega}^\text{II}v_{\omega}^{R}
        +
        \delta_{-\Omega',\omega}^\text{II}\bar{v}_{\omega}^{R}
        \big),
        \\
        \hat{a}_{-\Omega'}^\text{I}
        &=
        \int_{0}^{\infty}d\omega
        \big(
        \bar{\gamma}_{-\Omega',\omega}^\text{I}\hat{b}_{\omega}^{R}
        -
        \bar{\delta}_{-\Omega',\omega}^\text{I}(\hat{b}_{\omega}^{R})^{\dagger}
        \big),
        \\
        \hat{a}_{-\Omega'}^\text{II}
        &=
        \int_{0}^{\infty}d\omega
        \big(
        \bar{\gamma}_{-\Omega',\omega}^\text{II}\hat{b}_{\omega}^{R}
        -
        \bar{\delta}_{-\Omega',\omega}^\text{II}(\hat{b}_{\omega}^{R})^{\dagger}
        \big),
    \end{aligned}
\end{equation}
where $\{\gamma_{-\Omega',\omega}^\text{I},\delta_{-\Omega',\omega}^\text{I},\gamma_{-\Omega',\omega}^\text{II},\delta_{-\Omega',\omega}^\text{II}\}$ are the Bogoliubov coefficients that carry the information of the mirror's history.

To this end, the frequency spectrum of the \textit{out} particles observed by the inertial observer at $\mathscr{I}_{R}^{+}$ can be decomposed into the Rindler-Milne pair by (using Appendix~\ref{A1})
\begin{equation}\label{spectrum-squeezed-2}
    \begin{aligned}
        &\langle 0;in|
        (\hat{b}_{\omega}^{R})^{\dagger}\hat{b}_{\omega}^{R}
        |0;in\rangle
        \\
        &=
        \int_{0}^{\infty}d\Omega'
        \Biggl\{
        \frac{|\gamma_{-\Omega',\omega}^\text{I}|^{2}}{e^{\frac{2\pi\Omega'}{\kappa}}-1}
        \left(
        1+
        e^{\frac{2\pi\Omega'}{\kappa}}
        \frac{|\delta_{-\Omega',\omega}^\text{I}|^{2}}{|\gamma_{-\Omega',\omega}^\text{I}|^{2}}
        \right)  \\
        &\qquad\qquad\qquad +
        \frac{|\gamma_{-\Omega',\omega}^\text{II}|^{2}}{e^{\frac{2\pi\Omega'}{\kappa}}-1}
        \left(
        1+
        e^{\frac{2\pi\Omega'}{\kappa}}
        \frac{|\delta_{-\Omega',\omega}^\text{II}|^{2}}{|\gamma_{-\Omega',\omega}^\text{II}|^{2}}
        \right)
        +
        \frac{2\mathrm{Re}\big(\gamma_{-\Omega',\omega}^\text{I}\delta_{-\Omega',\omega}^\text{II}+\delta_{-\Omega',\omega}^\text{I}\gamma_{-\Omega',\omega}^\text{II} \big)}{e^{\frac{\pi\Omega'}{\kappa}}-e^{-\frac{\pi\Omega'}{\kappa}}}
        \Biggr\}.
    \end{aligned}
\end{equation}

This shows that the right-moving particles observed by the inertial \textit{out} observer at $\mathscr{I}_{R}^{+}$ are now manifestations of (i) individual two-mode squeezing of Rindler modes (the first term), (ii) individual two-mode squeezing of Milne modes (the second term), and (iii) mutual two-mode squeezing between Rindler and Milne modes (the third term). Therefore, the frequency $\omega$-spectrum generally deviates from the Bose--Einstein distribution, which is the behavior of Rindler/Milne modes. Similar to the discussion in the previous section, the information of the entire history of the mirror's trajectory is encoded in the Bogoliubov coefficients: $\{\gamma_{-\Omega',\omega}^\text{I},\delta_{-\Omega',\omega}^\text{I},\gamma_{-\Omega',\omega}^\text{II},\delta_{-\Omega',\omega}^\text{II}\}$. The above expression cleanly separates the Bose--Einstein behavior from the squeezing distortions; it also explicitly involves the notions of Milne (Hawking) modes and Rindler (partner) modes.

\subsubsection{Correlations measured at $\mathscr{I}_{R}^{+}$}

The trajectory considered in this section is timelike both in the past and future, so the radiated particles can only propagate to the mirror's right-hand side. In this case, the particles observed by an inertial observer at $\mathscr{I}_{R}^{+}$ consist of both squeezed Rindler and Milne modes. Therefore, the correlation between different modes of the radiated $\textit{out}$ particles becomes more complicated not only due to the inclusion of the Rindler mode (\textit{c.f.} there is only Milne mode in the previous asymptotically null trajectory), but also due to the individual/mutual squeezing effects. 

According to Eq.~\eqref{Q-perfect-non-asy}, the correlation between the right-moving $\textit{out}$ particles is determined by $\{\alpha_{\omega,-\omega'}^{R},\beta_{\omega,-\omega'}^{R}\}$, which can be expressed in terms of $\{\tilde{\alpha}_{-\Omega',-\omega'}^\text{I},\tilde{\beta}_{-\Omega',-\omega'}^\text{I},\tilde{\alpha}_{-\Omega',-\omega'}^\text{II},\tilde{\beta}_{-\Omega',-\omega'}^\text{II}\}$ or simply $\{\tilde{\alpha}_{-\Omega',-\omega'}^\text{II},\tilde{\beta}_{-\Omega',-\omega'}^\text{II}\}$ by
\begin{equation}
    \begin{aligned}
        \beta_{\omega,-\omega'}^{R}
        &=
        -
        \frac{1}{2\pi}
        \sqrt{\frac{\omega'}{\omega}}
        \int_{-\infty}^{\infty}dv\,
        e^{-i\omega' v}e^{-i\omega F(v)}
        \\
        &=
        \int_{0}^{\infty}d\Omega'
        \big(
        A_{-\Omega',\omega}^\text{I,II}
        \tilde{\beta}_{-\Omega',-\omega'}^\text{II}
        +
        B_{-\Omega',\omega}^\text{I,II}
        \bar{\tilde{\alpha}}_{-\Omega',-\omega'}^\text{II}
        \big),
        \\
        \alpha_{\omega,-\omega'}^{R}
        &=
        -
        \frac{1}{2\pi}
        \sqrt{\frac{\omega'}{\omega}}
        \int_{-\infty}^{\infty}dv\,
        e^{i\omega' v}e^{-i\omega F(v)}
        \\
        &=
        \int_{0}^{\infty}d\Omega'
        \big(
        C_{-\Omega',\omega}^\text{I,II}
        \tilde{\alpha}_{-\Omega',-\omega'}^\text{II}
        +
        D_{-\Omega',\omega}^\text{I,II}
        \bar{\tilde{\beta}}_{-\Omega',-\omega'}^\text{II}
        \big),
    \end{aligned}
\end{equation}
where Appendix~\ref{A3} is used, and the information about the mirror's history is encoded in the coefficients:
\begin{equation}
    \begin{aligned}
        A_{-\Omega',\omega}^\text{I,II}
        &:=
        \bar{\gamma}_{-\Omega',\omega}^\text{II}
        +
        e^{\frac{\pi\Omega'}{\kappa}}\delta_{-\Omega',\omega}^\text{I},
        \\
        B_{-\Omega',\omega}^\text{I,II}
        &:=
        -\big(
        \delta_{-\Omega',\omega}^\text{II}
        +
        e^{-\frac{\pi\Omega'}{\kappa}}
        \bar{\gamma}_{-\Omega',\omega}^\text{I}
        \big),
        \\
        C_{-\Omega',\omega}^\text{I,II}
        &:=
        \bar{\gamma}_{-\Omega',\omega}^\text{II}
        +
        e^{-\frac{\pi\Omega'}{\kappa}}\delta_{-\Omega',\omega}^\text{I},
        \\
        D_{-\Omega',\omega}^\text{I,II}
        &:=
        -\big(
        \delta_{-\Omega',\omega}^\text{II}
        +
        e^{\frac{\pi\Omega'}{\kappa}}
        \bar{\gamma}_{-\Omega',\omega}^\text{I}
        \big),
    \end{aligned}
\end{equation}
where the appearance of the factors $e^{\pm\frac{\pi\Omega'}{\kappa}}$ here is originated from the entanglement between the Rindler/Milne modes.

The correlation $Q^{RR}(\omega_{1},\omega_{2})$ between the right-moving \textit{out} particles according to Eq.~\eqref{Q-perfect-non-asy} is, therefore,
\begin{equation}\label{correlation-squeezed-2}
    \begin{aligned}
        &Q^{RR}(\omega_{1},\omega_{2})
        \\
        &=
        \int_{0}^{\infty}d\Omega_{1}'
        \frac{e^{-\frac{\pi\Omega_{1}'}{\kappa}}A_{-\Omega_{1}',\omega_{1}}^\text{I,II}\bar{A}_{-\Omega_{1}',\omega_{2}}^\text{I,II}+e^{\frac{\pi\Omega_{1}'}{\kappa}}B_{-\Omega_{1}',\omega_{1}}^\text{I,II}\bar{B}_{-\Omega_{1}',\omega_{2}}^\text{I,II} }{2\sinh(\pi\Omega_{1}'/\kappa)}
        \int_{0}^{\infty}d\Omega_{2}'
        \frac{e^{\frac{\pi\Omega_{2}'}{\kappa}}\bar{C}_{-\Omega_{2}',\omega_{1}}^\text{I,II}C_{-\Omega_{2}',\omega_{2}}^\text{I,II}+e^{-\frac{\pi\Omega_{2}'}{\kappa}}\bar{D}_{-\Omega_{2}',\omega_{1}}^\text{I,II}D_{-\Omega_{2}',\omega_{2}}^\text{I,II} }{2\sinh(\pi\Omega_{2}'/\kappa)}
        \\
        &+
        \int_{0}^{\infty}d\Omega_{1}'
        \frac{e^{-\frac{\pi\Omega_{1}'}{\kappa}}A_{-\Omega_{1}',\omega_{1}}^\text{I,II}D_{-\Omega_{1}',\omega_{2}}^\text{I,II}+e^{\frac{\pi\Omega_{1}'}{\kappa}}B_{-\Omega_{1}',\omega_{1}}^\text{I,II}C_{-\Omega_{1}',\omega_{2}}^\text{I,II} }{2\sinh(\pi\Omega_{1}'/\kappa)}
        \int_{0}^{\infty}d\Omega_{2}'
        \frac{e^{\frac{\pi\Omega_{2}'}{\kappa}}\bar{C}_{-\Omega_{2}',\omega_{1}}^\text{I,II}\bar{B}_{-\Omega_{2}',\omega_{2}}^\text{I,II}+e^{-\frac{\pi\Omega_{2}'}{\kappa}}\bar{D}_{-\Omega_{2}',\omega_{1}}^\text{I,II}\bar{A}_{-\Omega_{2}',\omega_{2}}^\text{I,II} }{2\sinh(\pi\Omega_{2}'/\kappa)},
    \end{aligned}
\end{equation}
where, as in Eq.~\eqref{correlation-squeezed-1}, the Boltzmann factors $e^{\pm\frac{\pi\Omega_{1,2}'}{\kappa}}$ in front of the coefficients $\{A^{\mathrm{I,II}},B^{\mathrm{I,II}},C^{\mathrm{I,II}},D^{\mathrm{I,II}}\}$ are originated from the transformation between the \textit{in} mode and the Rindler/Milne modes (see Eq.~\eqref{bogo-tilde}), whereas the same Boltzmann factors in the coefficients are originated from the entanglement between the Rindler/Milne modes.

\section{Discussions and summary }
\label{Discussions}

In the previous sections, we first reviewed the moving mirror model in the standard \textit{in-out} formulation, and then reformulated it in terms of Rindler/Milne modes to make the notion of entangled partner modes explicit, which also permits the explanation of the deviation of the radiation spectrum from the Bose--Einstein distribution observed by an inertial observer at $\mathscr{I}_{R}^{+}$. For timelike to null trajectory, the deviation is due to the two-mode squeezing of Milne modes (see Eq.~\eqref{spectrum-squeezed-1}); for timelike to timelike trajectory, it is due to the individual two-mode squeezing of the Milne/Rindler modes, and also their mutual two-mode squeezing (see Eq.~\eqref{spectrum-squeezed-2}). In these two situations, the squeezing caused by the mirror-field interaction generates additional quantum correlations between the radiated modes (see Eqs.~\eqref{correlation-squeezed-1} $\&$ \eqref{correlation-squeezed-2}). In this section, we demonstrate these deviations explicitly.

\subsection{Spectrum deviation and additional correlations}

\begin{figure}[h]
\centering
\subfloat[Green: Mirror never stops accelerating after $T=0$. Red, Blue: Mirror eventually stops accelerating.%
\label{fig-trajectory}]
{
\includegraphics[width=0.45\linewidth]{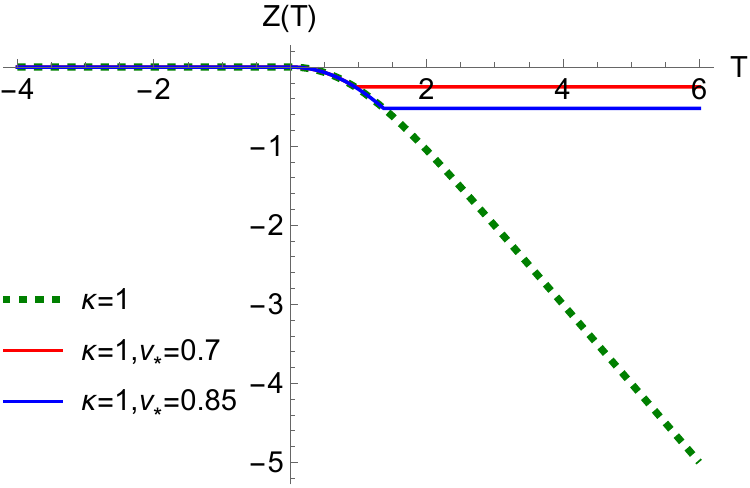}
}
\hfill
\subfloat[Green: Mirror never stops accelerating and asymptotes $v_{0}=1$. Red, Blue: Mirror eventually stops accelerating.%
\label{fig-ray-tracing}]
{
\includegraphics[width=0.45\linewidth]{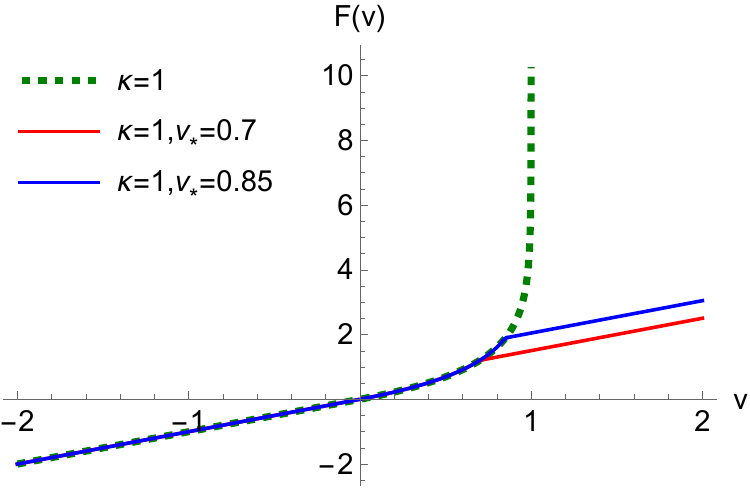}
}
\caption{Motions of the moving mirror with asymptotically null \eqref{traj-1} (green dashed) and with the asymptotically time-like motion \eqref{traj-2} (red, blue) (left) and the corresponding ray-tracing functions \eqref{raytracing-1-F} and \eqref{raytracing-2-F} (right).}
\label{fig-combined1}
\end{figure}

The timelike to null trajectory \eqref{traj-1} considered in Sect.~\ref{Formation of eternal black holes} is plotted as the green dashed curve in Fig.~\ref{fig-trajectory}, and the timelike to timelike trajectory \eqref{traj-2} in Sect.~\ref{Formation and evaporation of black holes} is plotted as the red/blue curves with different stopping parameter $v_{*}$ in Fig.~\ref{fig-trajectory}. The ray-tracing function corresponding to trajectory \eqref{traj-1} is plotted as the green dashed curve in Fig.~\ref{fig-ray-tracing}, and the ray-tracing function corresponding to trajectory \eqref{traj-2} is plotted as the red/blue curves in Fig.~\ref{fig-ray-tracing}. By using Eq.~\eqref{bogo-conventional}, we plot the corresponding frequency spectrum $|\beta_{\omega,-\omega'}^{R}|^{2}$ in Fig.~\ref{fig-combined2}.

\subsubsection{Timelike to asymptotically null trajectory}
\label{Timelike to asymptotically null trajectory}

\begin{figure}[h]
\centering
\subfloat[Green: Timelike to asymptotically null motion \eqref{traj-1}. Red, Blue: Timelike to timelike motion \eqref{traj-2}.%
\label{fig-beta1}]
{
\includegraphics[width=0.45\linewidth]{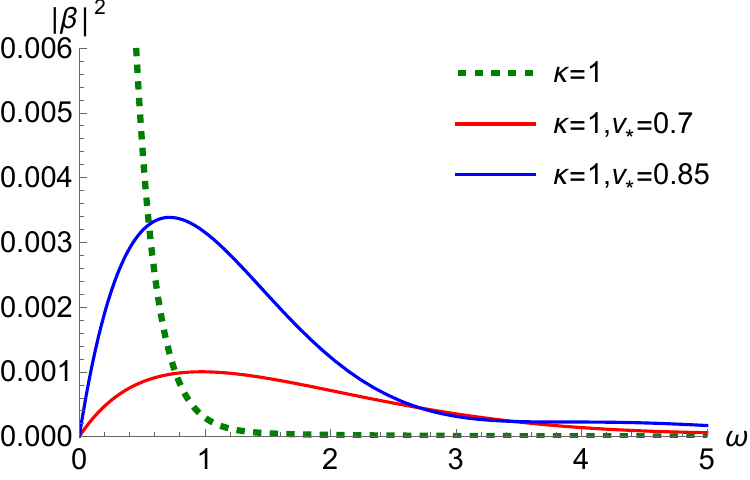}
}
\hfill
\subfloat[Green: Timelike to asymptotically null motion \eqref{traj-1}. Red, Blue: Timelike to timelike motion \eqref{traj-2}.%
\label{fig-beta2}]
{
\includegraphics[width=0.45\linewidth]{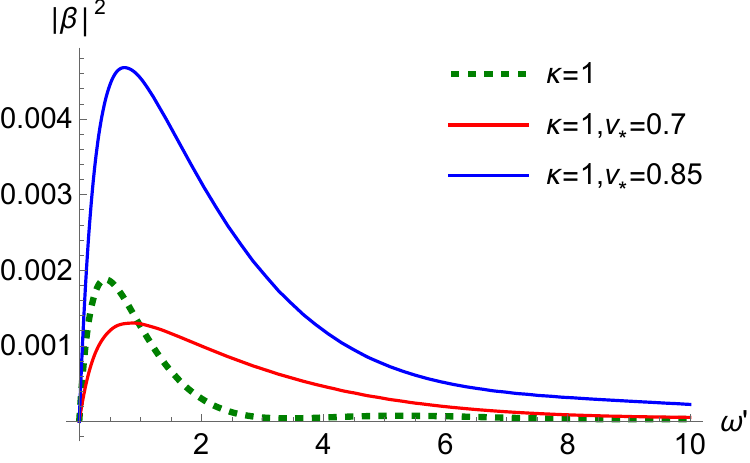}
}
\caption{Frequency spectra of the right-moving radiation $|\beta_{\omega,-\omega'}^R|^2$ at a fixed \textit{in} mode frequency $\omega'=2$ (left), and at a fixed \textit{out} mode frequency $\omega=1$ (right). According to Eq.~\eqref{beta-perfect-traj-1}, the timelike to asymptotically null spectrum $|\beta_{\omega,-\omega'}^{R}|^{2}$ diverges when $\omega=0$ at a given $\omega'\neq 0$ since $|\beta_{\omega,-\omega'}^{R}|^{2}\sim 1/\omega$, while $|\beta_{\omega,-\omega'}^{R}|^{2}=0$ when $\omega'=0$ at a given $\omega \neq 0$ since $|\beta_{\omega,-\omega'}^{R}|^{2}\sim \omega'$ ($c.f.$ $|\beta_{\omega,-\omega'}^{R}|^{2}\propto 1/\omega'$ for the Carlitz-Willey trajectory \eqref{bogo-ideal}). According to Eq.~\eqref{beta-2}, the timelike to timelike spectrum $|\beta_{\omega,-\omega'}^{R}|^{2}=0$ when $\omega=0$ at a given $\omega'\neq 0$ since $|\beta_{\omega,-\omega'}^{R}|^{2}\sim \omega$, and $|\beta_{\omega,-\omega'}^{R}|^{2}=0$ when $\omega'=0$ at a given $\omega \neq 0$ since $|\beta_{\omega,-\omega'}^{R}|^{2}\sim \omega'$.  }
\label{fig-combined2}
\end{figure}

According to Eq.~\eqref{beta-perfect-traj-1} for a timelike to asymptotically null trajectory, the spectrum $|\beta_{\omega,-\omega'}^{R}|^{2}$ diverges at $\omega=0$ when $\omega'\neq 0$ due to the $1/\omega$ behavior (see the green dashed curve in Fig.~\ref{fig-beta1}). However, the appearance of a divergence may not be a surprise. For example, in the Carlitz-Willey trajectory, the spectrum \eqref{bogo-ideal} also diverges at $\omega=0$ when $\omega'\neq 0$ due to the (1+1)-dimensional Bose--Einstein distribution. In higher spacetime dimensions, the Bose--Einstein distribution involves additional powers of $\omega$ in the numerator due to the transverse momentum, making the spectrum vanish instead at $\omega=0$. Alternatively, when considering particle detectors, there will also be a lower frequency cutoff due to the limitation of detections \cite{PhysRevD.88.025023}. Furthermore, in reality, the reflectivity of a mirror depends on the frequency, i.e., the high reflectivity behavior should appear only for sufficiently high frequencies. Therefore, there is also a lower frequency cutoff, below which the perfect reflectivity assumption fails \cite{Lin2021}. In contrast, $|\beta_{\omega,-\omega'}^{R}|^{2}$ for Eq.~\eqref{beta-perfect-traj-1} vanishes at $\omega'=0$ (see the green dashed curve in Fig.~\ref{fig-beta2}), instead of diverging like Eq.~\eqref{bogo-ideal} due to eternal acceleration \cite{PhysRevD.36.2327}.

\begin{figure}[h]
\centering
\subfloat[Correlation function $Q^{RR}(\omega_{1},\omega_{2})$ according to Eq.~\eqref{Q-perfect-asy} for timelike to asymptotically null motion \eqref{traj-1}.%
\label{fig-QRR-null}]
{
\includegraphics[width=0.45\linewidth]{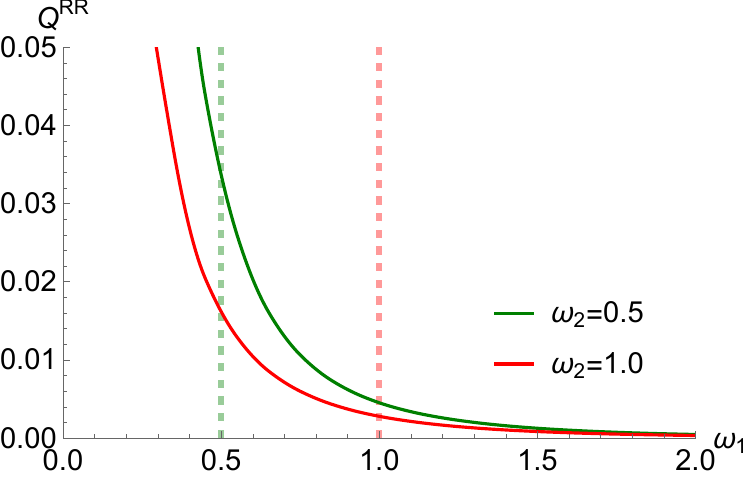}
}
\hfill
\subfloat[Correlation function $Q^{RL}(\omega_{1},-\omega_{2})$ according to Eq.~\eqref{Q-perfect-asy} for timelike to asymptotically null motion \eqref{traj-1}.%
\label{fig-QRL-null}]
{
\includegraphics[width=0.45\linewidth]{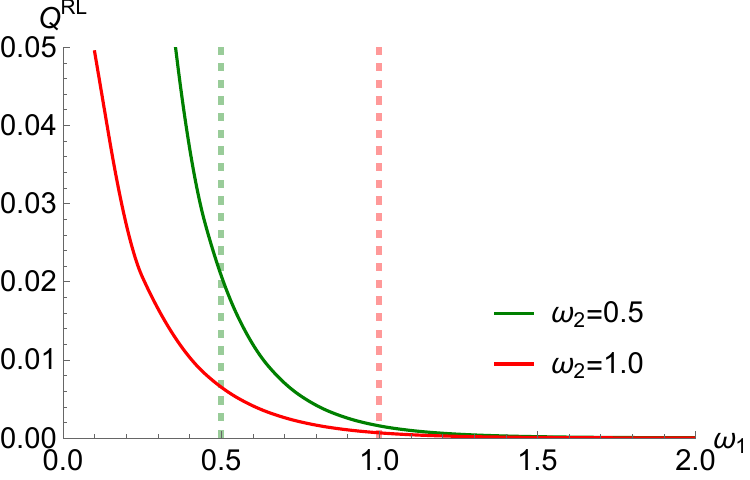}
}
\caption{
When the Milne modes are squeezed, the radiation received at $\mathscr{I}_{R}^{+}$ develops non-vanishing correlations. In the case of timelike to asymptotically null mirror trajectory, the correlation functions monotonically decreases as the \textit{out} mode frequencies $\omega_{1,2}$ increase, behaving like the spectrum (green dashed) in Fig.~\ref{fig-beta1}. (The vertical dashed lines indicate the locations of the chosen values of $\omega_{2}$.)
}   
\label{fig-combined3}
\end{figure}

Figure \ref{fig-combined3} illustrates the correlation functions $Q^{RR}(\omega_{1},\omega_{2})$ and $Q^{RL}(\omega_{1},-\omega_{2})$ for the timelike to asymptotically null trajectory \eqref{traj-1}. Recall that, in the case of Carlitz-Willey trajectory, the right-moving Minkowski plane wave \textit{out} modes and the left-moving Rindler \textit{out} modes are both self-uncorrelated, i.e., $Q^{RR}(\omega_{1},\omega_{2})=Q^{LL}(-\omega_{1},-\omega_{2})=0$ in Eq.~\eqref{Q-0}, while the correlation between the right-moving Minkowski plane wave \textit{out} modes and the left-moving Rindler \textit{out} modes is strongly peak, i.e., $Q^{RL}(\omega_{1},-\omega_{2})\propto [\delta(\omega_{1}-\omega_{2})]^{2}$ in Eq.~\eqref{Q-delta}. Therefore, Fig.~\ref{fig-combined3} shows that the two-mode squeezing  
of the right-moving reflected Milne modes by trajectory \eqref{traj-1} results in additional correlation between the right-moving Minkowski plane wave \textit{out} modes so that $Q^{RR}(\omega_{1},\omega_{2})\neq 0$, which further alters the correlation between the right-moving Minkowski plane wave \textit{out} modes and the left-moving Rindler \textit{out} modes so that $Q^{RL}(\omega_{1},-\omega_{2})\neq 0$ when $\omega_{1}\neq\omega_{2}$. However, both $Q^{RR}(\omega_{1},\omega_{2})$ and $Q^{RL}(\omega_{1},-\omega_{2})$ turn out to suffer from the infrared divergence due to the divergence of the Bogoliubov $\beta$-coefficient mentioned in the last paragraph.

\subsubsection{Timelike to timelike trajectory}
\label{Timelike to timelike trajectory}

In the case of timelike to timelike trajectory \eqref{traj-2}, the spectrum $|\beta_{\omega,-\omega'}^{R}|^{2}$ according to Eq.~\eqref{beta-2} is plotted as the blue and red curves in Fig.~\ref{fig-combined2}, which show that the spectra are finite everywhere in contrast to the green dashed curve in Fig.~\ref{fig-beta1}, and the longer the mirror accelerates (larger $v_{*}$ but still $v_{*}<v_{0}$), the more particles are created. We named a few examples that are able to make the spectrum finite at $\omega=0$ in the previous subsection \ref{Timelike to asymptotically null trajectory}. Here, by returning the mirror to rest provides an alternative mean to achieve this. In addition, since $|\beta_{\omega,-\omega'}^{R}|^{2}\sim \omega$ according to Eq.~\eqref{beta-2} at small $\omega$, the spectrum in fact vanishes at $\omega=0$. This indicates that the spectrum will have a peak in the intermediate $\omega$ regime, since the spectrum eventually decays at large $\omega$.

\begin{figure}[h]
\centering
\subfloat[$Q^{RR}(\omega_{1},\omega_{2})$ according to Eq.~\eqref{Q-perfect-non-asy} for timelike to timelike motion \eqref{traj-2} with $\kappa=1$ and $v_{*}=0.7$.%
\label{fig-QRR-all}]
{
\includegraphics[width=0.45\linewidth]{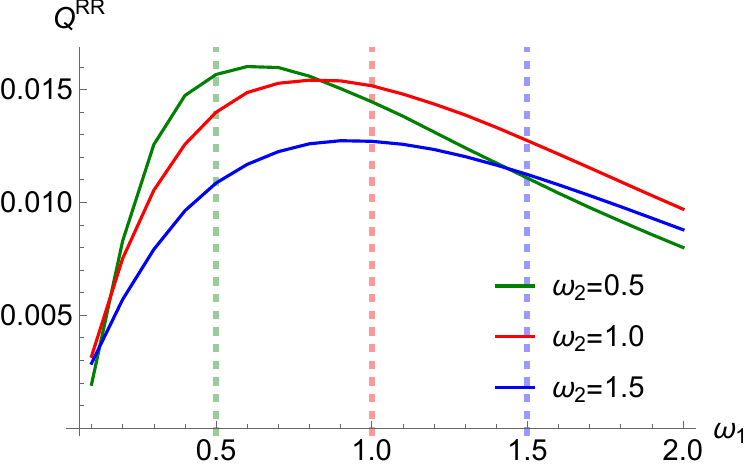}
}
\hfill
\subfloat[$Q^{RR}(\omega_{1},\omega_{2})$ according to Eq.~\eqref{Q-perfect-non-asy} for timelike to timelike motion \eqref{traj-2} with $\kappa=1$.%
\label{fig-QRR-all2}]
{
\includegraphics[width=0.45\linewidth]{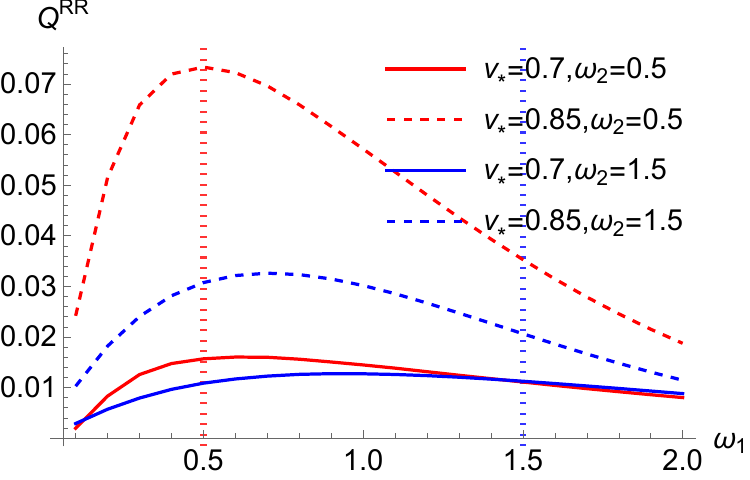}
}
\caption{The mirror that begins and ends at the timelike infinities makes the past and future null infinities as valid Cauchy surfaces. Thus, the squeezing of the Rindler/Milne modes results in non-vanishing correlation in the radiation received at $\mathscr{I}_{R}^{+}$. However, instead of monotonically decreasing, the correlations exhibit local maxima, which generally do not occur at exactly $\omega_{1}=\omega_{2}$, according to the distribution of the Bogoliubov coefficient (see Fig.~\ref{fig-beta1}). The left panel illustrates that, the larger the Minkowski plane wave $out$ mode $\omega_{2}$ is measured, the farther away it is from its mostly correlated mode $\omega_{1}$. The right panel shows that, the longer the mirror undergoes acceleration, the larger the correlation. (The vertical dashed lines indicate the locations of the chosen values of $\omega_{2}$.)  } 
\label{fig-combined4}
\end{figure}

By using Eq.~\eqref{Q-perfect-non-asy}, the correlation functions $Q^{RR}(\omega_{1},\omega_{2})$ for trajectory \eqref{traj-2} are plotted in Fig.~\ref{fig-combined4}, which behave similarly to the timelike to timelike spectrum in Fig.~\ref{fig-combined2} with a frequency peak. In fact, the correlation functions in Fig.~\ref{fig-combined3} also behave similarly to the timelike to asymptotically null spectrum in Fig.~\ref{fig-combined2}. By inspecting the expressions in the correlations \eqref{Q-perfect-asy} and \eqref{Q-perfect-non-asy}, the similarities may be originated from the sharp peak behavior due to the static motion contribution in the Bogoliubov $\alpha$-coefficients in Eqs.~\eqref{alpha-perfect-traj-1} and \eqref{alpha-2}. Similar to a Dirac delta distribution, this makes the behavior of the products of the second terms in $\{Q^{RR},Q^{RL}\}$ in Eqs.~\eqref{Q-perfect-asy} and \eqref{Q-perfect-non-asy} to be governed by the Bogoliubov $\beta$-coefficients, resulting in the similarity between the correlation functions and frequency spectra.

Figure \ref{fig-combined4} also shows that the peak of the correlation does not occur at the same Minkowski frequency $\omega_{1}=\omega_{2}$. Instead, the larger the frequency $\omega_{2}$ is measured, the farther away it is from the mostly correlated frequency $\omega_{1}$. However, as given by Eq.~\eqref{correlation-squeezed-2}, the content of $Q^{RR}(\omega_{1},\omega_{2})$ is in fact a complex combination of various mode squeezing effects, so it is difficult at this point to provide a precise explanation for this tendency.

\subsection{Entanglement entropy vs. quantum correlation}

Black hole information paradox is one of the biggest issues that remains to be resolved in modern physics. Essentially, this concerns the fate of the partner mode \cite{PhysRevD.91.124060,PhysRevD.101.024003,PhysRevD.110.025023} of the Hawking mode, which requires analyses that go beyond the semi-classical evaporation of a black hole. Since we currently lack direct access to observe the quantum effects of a black hole, analog models of black hole physics have been proposed \cite{Fulling1976,Davies1977,PhysRevLett.46.1351,PhysRevLett.85.4643,PhysRevD.105.105009}. This is supported by the (renormalized) time-dependent entanglement entropy defined by Bianchi and Smerlak \cite{PhysRevD.90.041904}, albeit it involves the notion of negative energy flux \cite{walker1985negative,PhysRevD.90.041904,PhysRevD.70.125008,GoodOng,Bianchi:2026xoi}.

In the moving mirror model with perfect reflectivity, the partner (Rindler) mode of the Hawking (Milne) mode is guaranteed to eventually become an accessible degree of freedom for the inertial observer at $\mathscr{I}_{R}^{+}$, in addition to the previously received Hawking mode. The time-dependent entanglement entropy then follows the so-called \textit{Page curve} \cite{page93a,page93b}, which is proposed to illustrate the unitary evolution of black hole evaporation. However, the relation between the time-dependent entanglement entropy and the time-independent vacuum state that explicitly involves the notion of partner modes developed in this paper is not clear.

In addition, since the time-dependent entanglement entropy and the energy flux are generally divergent, the renormalization process typically involves subtracting the contribution from vacuum fluctuations \cite{Fulling1976}. Meanwhile, since the Rindler/Milne-mode pair is regarded as vacuum fluctuations in the \textit{in} vacuum, there may be some subtleties regarding the renormalization of the entropy and energy flux \cite{PhysRevD.110.025023}.

Nevertheless, the squeezing effects discussed in this paper shall not affect the entanglement entropy. For example, in the case of timelike to asymptotically null mirror, the density matrix of the total system is given by
\begin{equation}
    \begin{aligned}
        \hat{\rho}_{\mathrm{I,II}}
        &:=
        |0_{-\Omega'};in\rangle\langle 0_{-\Omega'};in|
        \\
        &=
        \frac{1}{|\tilde{\alpha}_{-\Omega'}|^{2}}\sum_{n,n'=0}^{\infty}
        \left(
        -\frac{\bar{\tilde{\beta}}_{-\Omega'}}{\tilde{\alpha}_{-\Omega'}}
        \right)^{n}
        \left(
        -\frac{\tilde{\beta}_{-\Omega'}}{\bar{\tilde{\alpha}}_{-\Omega'}}
        \right)^{n'}
        \big(|n_{-\Omega'}\rangle_{\mathrm{I}}
        |n_{-\Omega'}\rangle_{\mathrm{II}}\big)
        \big({}_{\mathrm{I}}\langle n'_{-\Omega'}|
        {}_{\mathrm{II}}\langle n'_{-\Omega'}|\big)
        \\
        &=
        \hat{f}_{\mathrm{I}}
        \hat{f}_{\mathrm{II}}
        \left(
        \frac{1}{|\tilde{\alpha}_{-\Omega'}|^{2}}\sum_{n,n'=0}^{\infty}
        \left(
        -\frac{\bar{\tilde{\beta}}_{-\Omega'}}{\tilde{\alpha}_{-\Omega'}}
        \right)^{n}
        \left(
        -\frac{\tilde{\beta}_{-\Omega'}}{\bar{\tilde{\alpha}}_{-\Omega'}}
        \right)^{n'}
        |n_{-\Omega'}\rangle_{L}
        |n_{-\Omega'}\rangle_{R}
        {}_{L}\langle n'_{-\Omega'}|
        {}_{R}\langle n'_{-\Omega'}|
        \right)
        \hat{f}_{\mathrm{II}}^{\dagger}
        \hat{f}_{\mathrm{I}}^{\dagger}
        \\
        &:=
        \hat{f}_{\mathrm{I}}
        \hat{f}_{\mathrm{II}}
        \hat{\rho}_{LR}
        \hat{f}_{\mathrm{II}}^{\dagger}
        \hat{f}_{\mathrm{I}}^{\dagger}.
    \end{aligned}
\end{equation}
where the individual two-mode squeezing operators $\hat{f}_{\mathrm{I,II}}$ are given by Eq.~\eqref{fIfII-null}. The entanglement entropy between the Rindler mode and the Milne mode can be determined by the reduced density matrix of the Rindler mode:
\begin{equation}
    \begin{aligned}
        \hat{\rho}_{\mathrm{I}}
        &:=
        \mathrm{Tr}_{\mathrm{II}}\hat{\rho}_{\mathrm{I,II}}
        =
        \frac{1}{|\tilde{\alpha}_{-\Omega'}|^{2}}\sum_{n}^{\infty}
        \frac{|\bar{\tilde{\beta}}_{-\Omega'}|^{2n}}{|\tilde{\alpha}_{-\Omega'}|^{2n}}
        \big(|n_{-\Omega'}\rangle_{\mathrm{I}}\big)
        \big({}_{\mathrm{I}}\langle n_{-\Omega'}|\big),
    \end{aligned}
\end{equation}
or the reduced density matrix of the Milne mode:
\begin{equation}
    \begin{aligned}
        \hat{\rho}_{\mathrm{II}}
        :=
        \mathrm{Tr}_{\mathrm{I}}\hat{\rho}_{\mathrm{I,II}}
        =
        \frac{1}{|\tilde{\alpha}_{-\Omega'}|^{2}}\sum_{n}^{\infty}
        \frac{|\bar{\tilde{\beta}}_{-\Omega'}|^{2n}}{|\tilde{\alpha}_{-\Omega'}|^{2n}}
        \big(|n_{-\Omega'}\rangle_{\mathrm{II}}\big)
        \big({}_{\mathrm{II}}\langle n_{-\Omega'}|\big).
    \end{aligned}
\end{equation}

Alternatively, in the perspective of the left-moving Rindler \textit{out} mode and the right-moving Minkowski plane wave \textit{out} mode, which sees the squeezed Milne modes, the reduced density matrix of the left-moving Rindler \textit{out} mode is
\begin{equation}
    \begin{aligned}
        \hat{\rho}_{L}
        &:=
        \mathrm{Tr}_{R}\hat{\rho}_{\mathrm{I,II}}
        =
        \mathrm{Tr}_{R}
        \left(
        \hat{f}_{\mathrm{I}}
        \hat{f}_{\mathrm{II}}
        \hat{\rho}_{LR}
        \hat{f}_{\mathrm{II}}^{\dagger}
        \hat{f}_{\mathrm{I}}^{\dagger}
        \right)
        =
        \mathrm{Tr}_{R}\hat{\rho}_{LR}
        =
        \frac{1}{|\tilde{\alpha}_{-\Omega'}|^{2}}\sum_{n}^{\infty}
        \frac{|\bar{\tilde{\beta}}_{-\Omega'}|^{2n}}{|\tilde{\alpha}_{-\Omega'}|^{2n}}
        \big(|n_{-\Omega'}\rangle_{L}\big)
        \big({}_{L}\langle n_{-\Omega'}|\big),
    \end{aligned}
\end{equation}
and the reduced density matrix of the right-moving Minkowski plane wave \textit{out} mode is
\begin{equation}
    \begin{aligned}
        \hat{\rho}_{R}
        &:=
        \mathrm{Tr}_{L}\hat{\rho}_{\mathrm{I,II}}
        =
        \mathrm{Tr}_{L}
        \left(
        \hat{f}_{\mathrm{I}}
        \hat{f}_{\mathrm{II}}
        \hat{\rho}_{LR}
        \hat{f}_{\mathrm{II}}^{\dagger}
        \hat{f}_{\mathrm{I}}^{\dagger}
        \right)
        =
        \mathrm{Tr}_{L}\hat{\rho}_{LR}
        =
        \frac{1}{|\tilde{\alpha}_{-\Omega'}|^{2}}\sum_{n}^{\infty}
        \frac{|\bar{\tilde{\beta}}_{-\Omega'}|^{2n}}{|\tilde{\alpha}_{-\Omega'}|^{2n}}
        \big(|n_{-\Omega'}\rangle_{R}\big)
        \big({}_{R}\langle n_{-\Omega'}|\big),
    \end{aligned}
\end{equation}
where we have used the trace identity: $\mathrm{Tr}(ABC)=\mathrm{Tr}(BCA)=\mathrm{Tr}(CAB)$ so that the local unitary squeezing operators $\hat{f}_{\mathrm{I,II}}$ are canceled. Since $\{\hat{\rho}_{\mathrm{I}},\hat{\rho}_{\mathrm{II}}\}$ and $\{\hat{\rho}_{L},\hat{\rho}_{R}\}$ have identical eigenvalues, the entanglement entropy $(S_{\mathrm{vN}})$ between the left-moving Rindler \textit{out} mode and the right-moving Minkowski plane wave \textit{out} mode is identical to that between the standard left-moving Rindler mode and Milne mode:
\begin{equation}
    \begin{aligned}
        S_{\mathrm{vN}}(\Omega')
        &=
        \ln|\tilde{\alpha}_{-\Omega'}|^{2}
        -
        |\tilde{\beta}_{-\Omega'}|^{2}
        \ln|\tilde{\beta}_{-\Omega'}/\tilde{\alpha}_{-\Omega'}|^{2}.
    \end{aligned}
\end{equation}

Therefore, if an experiment measures the entanglement directly, then the squeezing effect is not important. However, if the experiment measures certain quantum correlations and attempts to reconstruct the entanglement entropy, then it is important to keep track of the squeezing effects in order to make appropriate interpretation about the measurement. Alternatively, perhaps by investigating the design of particle detectors, the issue can be better addressed \cite{6pz9-93h5}.


\subsection{Statistical inversion of radiation spectrum}

In this paper, we explained the deviation of the frequency $\omega$-spectrum of the radiated right-moving particles observed by an inertial observer at $\mathscr{I}_{R}^{+}$ from the Bose--Einstein distribution in the context of a perfectly reflecting mirror in (1+1)d by explicitly invoking the notions of Rindler mode and Milne mode. However, a more dramatic deviation of the spectrum is also known for a mirror with a non-trivial, i.e., time-dependent and frequency-dependent, reflectivity \cite{barton1995quantum}, which may change the spectrum from Bose--Einstein to Fermi--Dirac-like distribution \cite{PhysRevD.80.125003,Lin2020,Lin2021,PhysRevD.77.045011}.

In fact, other similar reversals of the statistical distribution of radiation have also been observed in the literature. For example, in a particular accelerated motion, an electric charge may radiate electromagnetic radiation with a Bose--Einstein-like distribution, while a scalar charge may radiate scalar radiation with a Fermi--Dirac-like distribution \cite{nikishov1995emission}. With these observations, it has been suggested that the statistical distribution of the radiation spectrum may not always reflect the algebra, i.e., commutation or anti-commutation relations, of the annihilation/creation operators of the relevant field operators \cite{PhysRevD.77.045011,Goodspin}. Instead, the details of the interaction between the charge source and the radiation affect the overall spectral distribution. Based on the discussions in this paper, we believe that these are also the squeezing effects in action.

Experimental measurements of such a statistical inversion, i.e., deviation of frequency spectrum from Bose--Einstein and becomes Fermi--Dirac-like albeit the quantum field obeys the commutation relation for a Boson, 
of the radiation spectrum may become possible in the future. It has been proposed to create a moving mirror with the desired motion to mimic black hole radiation using laser-plasma physics \cite{Chen2017prl,Chen2020,AnaBHEL,Liu2026}; however, this comes at the cost of the mirror achieving an extremely low reflectivity \cite{Liu2020}. In such a situation, one may imagine that the \textit{in} vacuum initially prepares entangled left-moving Rindler/Milne modes with Bose--Einstein distributions. However, after reflection of the Milne modes by the low-reflectivity timelike to asymptotically null mirror, the Milne modes are squeezed in such a way that the spectrum received at $\mathscr{I}_{R}^{+}$ becomes Fermi--Dirac-like. Moreover, the radiation received by an inertial observer at $\mathscr{I}_{R}^{+}$ will also involve contributions from the transmitted modes. The non-trivial reflectivity of the mirror will also demand a separate analysis of the correlation functions, which shall be important for experimental investigations of black hole information paradox. The analysis performed in the present work should provide the foundation for future investigations.

\subsection{Summary}

Finally, let us end with a brief summary of the main messages of this work below.

In the case of Carlitz-Willey trajectory, the spectrum $|\beta_{\omega,-\omega'}^{R}|^{2}$ is exactly Bose--Einstein \eqref{bogo-ideal} for all frequency $\omega$ since $\{\alpha_{\omega,-\Omega'}^\text{II},\beta_{\omega,-\Omega'}^\text{II}\}=\{\delta(\omega-\Omega'),0\}$ and $\beta_{\omega,-\omega'}^{R}=\tilde{\beta}_{-\omega,-\omega'}^\text{II}$ according to Eq.~\eqref{bogo-additional1}, meaning the absence of two-mode squeezing of Milne modes in Eq.~\eqref{spectrum-squeezed-1}. In addition, the correlation functions are given by $Q^{RR}(\omega_{1},\omega_{2})=Q^{LL}(-\omega_{1},-\omega_{2})=0$ in Eq.~\eqref{Q-0} and $Q^{RL}(\omega_{1},-\omega_{2})\propto [\delta(\omega_{1}-\omega_{2})]^{2}$ in Eq.~\eqref{Q-delta}. Therefore, in the \textit{absence of two-mode squeezing}, the radiated right-moving particles at $\mathscr{I}_{R}^{+}$ are \textit{exactly identical} to the self-uncorrelated Milne modes.

In the case of timelike to asymptotically null trajectory \eqref{traj-1}, the radiated right-moving particles at $\mathscr{I}_{R}^{+}$ are now the squeezed Milne modes instead according to Eq.~\eqref{spectrum-squeezed-1}. However, the squeezing is in such a way that the Bose--Einstein distribution can be approximated for certain frequency $\{\omega,\omega'\}$ regime, at the cost of inducing additional correlations discussed in the previous subsection \ref{Timelike to asymptotically null trajectory}. Therefore, these right-moving particles are \textit{Milne modes in disguise}.

In the case of timelike to timelike trajectory \eqref{traj-2}, the radiated right-moving particles at $\mathscr{I}_{R}^{+}$ are a combination of squeezed Milne modes, squeezed Rindler modes, and the information of their mutual entanglement according to Eq.~\eqref{spectrum-squeezed-2}. Therefore, the radiated particles at $\mathscr{I}_{R}^{+}$ are a complex \textit{mixture of everything}, introducing additional correlations discussed in the previous subsection \ref{Timelike to timelike trajectory} and preventing the recovery of Bose--Einstein distribution.

On the experimental side, it is important to realize that the measurement of quantum correlations, which are defined by the specific measurements in the laboratory, in general does not provide a direct measure of entanglement. Instead, the correlations may be contaminated by additional squeezing effects. This will be important for experiments aiming to investigate black hole information paradox in the lab.

Lastly, it is reported in the literature regarding the peculiar statistical inversion of the radiation spectrum compared to the (anti-)commutator algebra obeyed by the field operator in other systems \cite{PhysRevD.80.125003,Lin2020,Lin2021,PhysRevD.77.045011,nikishov1995emission,Goodspin}. It is suggested that the field-source interaction should be responsible for such phenomena. In this work, we demonstrated that this is indeed true and attributed such phenomena to the mode squeezing effects.

\begin{acknowledgments}
KNL was supported by an appointment to the YST fellow at the Asia Pacific Center for Theoretical Physics (APCTP) through the Korean Government's Science and Technology Promotion Fund and Lottery Fund. PC acknowledges the support of Taiwan’s National Science and Technology Council (NSTC), under the funding with account number 115-2112-M-002-014. Funding comes partly from the FY2024-SGP-1-STMM Faculty Development Competitive Research Grant (FDCRGP) no.201223FD8824 and SSH20224004 at Nazarbayev University in Qazaqstan. YN was partly supported by JSPS KAKENHI (Grant No.~JP23K25871).
\end{acknowledgments}

\appendix

\section{Useful integrals-1}
\label{A1}

\begin{equation}\label{integral-results}
    \begin{aligned}
        \int_{0}^{\infty}d\omega'
        \tilde{\beta}_{-\Omega_{1}',-\omega'}^\text{II}
        \bar{\tilde{\beta}}_{-\Omega_{2}',-\omega'}^\text{II}
        &=
        \frac{\kappa e^{-\frac{\pi(\Omega_{1}'+\Omega_{2}')}{2\kappa}}}{2\pi\sqrt{\Omega_{1}'\Omega_{2}'}}
        \Gamma\big(1+\frac{i\Omega_{1}'}{\kappa}\big)
        \Gamma\big(1-\frac{i\Omega_{2}'}{\kappa}\big)\delta(\Omega_{1}'-\Omega_{2}'),
        \\
        \int_{0}^{\infty}d\omega'
        \tilde{\beta}_{-\Omega_{1}',-\omega'}^\text{II}
        \tilde{\alpha}_{-\Omega_{2}',-\omega'}^\text{II}
        &=
        0,
        \\
        \int_{0}^{\infty}d\omega'
        \bar{\tilde{\alpha}}_{-\Omega_{1}',-\omega'}^\text{II}
        \bar{\tilde{\beta}}_{-\Omega_{2}',-\omega'}^\text{II}
        &=
        0,
        \\
        \int_{0}^{\infty}d\omega'
        \bar{\tilde{\alpha}}_{-\Omega_{1}',-\omega'}^\text{II}
        \tilde{\alpha}_{-\Omega_{2}',-\omega'}^\text{II}
        &=
        \frac{\kappa e^{\frac{\pi(\Omega_{1}'+\Omega_{2}')}{2\kappa}}}{2\pi\sqrt{\Omega_{1}'\Omega_{2}'}}
        \Gamma\big(1-\frac{i\Omega_{1}'}{\kappa}\big)
        \Gamma\big(1+\frac{i\Omega_{2}'}{\kappa}\big)\delta(\Omega_{1}'-\Omega_{2}'),
    \end{aligned}
\end{equation}

\section{Useful integrals-2}
\label{A2}

\begin{equation}
    \begin{aligned}
        \int_{-\infty}^{v_{0}}dv\,
        e^{i\omega' v}
        \big.u_{-\Omega'}^\text{II}\big|_{\mathscr{I}_{R}^{-}}
        &=
        -
        \frac{1}{\sqrt{4\pi\Omega'}}
        \int_{-\infty}^{v_{0}}dv\,
        e^{i\omega' v}
        e^{\frac{i\Omega'}{\kappa}\ln(1-\kappa v)}
        =
        \frac{2\pi}{\sqrt{4\pi\omega'}}\tilde{\alpha}_{-\Omega',-\omega'}^\text{II},
        \\
        \int_{-\infty}^{v_{0}}dv\,
        e^{i\omega' v}
        \big.\bar{u}_{-\Omega'}^\text{II}\big|_{\mathscr{I}_{R}^{-}}
        &=
        -
        \frac{1}{\sqrt{4\pi\Omega'}}
        \int_{-\infty}^{v_{0}}dv\,
        e^{i\omega' v}
        e^{-\frac{i\Omega'}{\kappa}\ln(1-\kappa v)}
        =
        \frac{2\pi}{\sqrt{4\pi\omega'}}\bar{\tilde{\beta}}_{-\Omega',-\omega'}^\text{II},
        \\
        \int_{-\infty}^{v_{0}}dv\,
        e^{-i\omega' v}
        \big.u_{-\Omega'}^\text{II}\big|_{\mathscr{I}_{R}^{-}}
        &=
        -
        \frac{1}{\sqrt{4\pi\Omega'}}
        \int_{-\infty}^{v_{0}}dv\,
        e^{-i\omega' v}
        e^{\frac{i\Omega'}{\kappa}\ln(1-\kappa v)}
        =
        \frac{2\pi}{\sqrt{4\pi\omega'}}\tilde{\beta}_{-\Omega',-\omega'}^\text{II},
        \\
        \int_{-\infty}^{v_{0}}dv\,
        e^{-i\omega' v}
        \big.\bar{u}_{-\Omega'}^\text{II}\big|_{\mathscr{I}_{R}^{-}}
        &=
        -
        \frac{1}{\sqrt{4\pi\Omega'}}
        \int_{-\infty}^{v_{0}}dv\,
        e^{-i\omega' v}
        e^{-\frac{i\Omega'}{\kappa}\ln(1-\kappa v)}
        =
        \frac{2\pi}{\sqrt{4\pi\omega'}}\bar{\tilde{\alpha}}_{-\Omega',-\omega'}^\text{II}.
    \end{aligned}
\end{equation}
\section{Useful integrals-3}
\label{A3}

\begin{equation}
    \begin{aligned}
        \int_{-\infty}^{\infty}dv\,
        e^{i\omega' v}
        \big.u_{-\Omega'}^\text{I}\big|_{\mathscr{I}_{R}^{-}}
        &=
        -
        \frac{1}{\sqrt{4\pi\Omega'}}
        \int_{v_{0}}^{\infty}dv\,
        e^{i\omega' v}
        e^{-\frac{i\Omega'}{\kappa}\ln(\kappa v-1)}
        =
        \frac{2\pi}{\sqrt{4\pi\omega'}}\tilde{\alpha}_{-\Omega',-\omega'}^\text{I},
        \\
        \int_{-\infty}^{\infty}dv\,
        e^{i\omega' v}
        \big.\bar{u}_{-\Omega'}^\text{I}\big|_{\mathscr{I}_{R}^{-}}
        &=
        -
        \frac{1}{\sqrt{4\pi\Omega'}}
        \int_{v_{0}}^{\infty}dv\,
        e^{i\omega' v}
        e^{\frac{i\Omega'}{\kappa}\ln(\kappa v-1)}
        =
        \frac{2\pi}{\sqrt{4\pi\omega'}}\bar{\tilde{\beta}}_{-\Omega',-\omega'}^\text{I},
        \\
        \int_{-\infty}^{\infty}dv\,
        e^{-i\omega' v}
        \big.u_{-\Omega'}^\text{I}\big|_{\mathscr{I}_{R}^{-}}
        &=
        -
        \frac{1}{\sqrt{4\pi\Omega'}}
        \int_{v_{0}}^{\infty}dv\,
        e^{-i\omega' v}
        e^{-\frac{i\Omega'}{\kappa}\ln(\kappa v-1)}
        =
        \frac{2\pi}{\sqrt{4\pi\omega'}}\tilde{\beta}_{-\Omega',-\omega'}^\text{I},
        \\
        \int_{-\infty}^{\infty}dv\,
        e^{-i\omega' v}
        \big.\bar{u}_{-\Omega'}^\text{I}\big|_{\mathscr{I}_{R}^{-}}
        &=
        -
        \frac{1}{\sqrt{4\pi\Omega'}}
        \int_{v_{0}}^{\infty}dv\,
        e^{-i\omega' v}
        e^{\frac{i\Omega'}{\kappa}\ln(\kappa v-1)}
        =
        \frac{2\pi}{\sqrt{4\pi\omega'}}\bar{\tilde{\alpha}}_{-\Omega',-\omega'}^\text{I}.
    \end{aligned}
\end{equation}
\nocite{*}

\bibliography{main}

\end{document}